\documentclass[%
preprint,
amsmath,amssymb,
aps,
prb,
]{revtex4-1}

\usepackage{array}
\usepackage{txfonts}
\usepackage{color}
\usepackage{booktabs}

\usepackage[pdftex]{graphicx}		% for arXiv

\usepackage{dcolumn}% Align table columns on decimal point
\usepackage{bm}% bold math
\begin{document}
	
	%\preprint{APS/123-QED}
	\title{Thermal Hall Effects of Spins and Phonons in Kagome Antiferromagnet Cd-Kapellasite}

	\author{Masatoshi Akazawa$^1$}
	\author{Masaaki Shimozawa$^{1,2}$}
	\author{Shunichiro Kittaka$^{1,3}$}
	\author{Toshiro Sakakibara$^1$}
	\author{Ryutaro Okuma$^{1,4}$}
	\author{Zenji Hiroi$^1$}
	\author{Hyun-Yong Lee$^{1,5}$}
	\author{Naoki Kawashima$^1$}
	\author{Jung Hoon Han$^{6}$}
	\author{Minoru Yamashita$^1$}
	\email[]{my@issp.u-tokyo.ac.jp}

	\affiliation{$^1$The Institute for Solid State Physics, The University of Tokyo, Kashiwa, 277-8581, Japan}
	\affiliation{$^2$Graduate School of Engineering Science, Osaka University, Toyonaka, 560-8531, Japan}
	\affiliation{$^3$Department of Physics, Chuo University, Kasuga, Bunkyo-ku, Tokyo 112-8551, Japan}
	\affiliation{$^4$Okinawa Institute of Science and Technology Graduate University, Kunigami-gun, 904-0495, Japan}
	\affiliation{$^5$Division of Display and Semiconductor Physics, Korea University, Sejong, 30019, Korea}
	\affiliation{$^6$Department of Physics, Sungkyunkwan University, Suwon 16419, Korea}

	\date{\today}% It is always \today, today,
	
	\begin{abstract}
		We have investigated the thermal-transport properties of the kagome antiferromagnet Cd-kapellasite (Cd-K).
		We find that a field suppression effect on the longitudinal thermal conductivity $\kappa_{xx}$ sets in below $\sim$25\,K. This field suppression effect at 15\,T becomes as large as 80\% at low temperatures, suggesting a large spin contribution $\kappa_{xx}^\textrm{sp}$ in $\kappa_{xx}$. 
		We also find clear thermal Hall signals in the spin liquid phase in all Cd-K samples.
		The magnitude of the thermal Hall conductivity $\kappa_{xy}$ shows a significant dependence on the sample's scattering time, as seen in the rise of the peak $\kappa_{xy}$ value in almost linear fashion with the magnitude of $\kappa_{xx}$. 
		On the other hand, the temperature dependence of $\kappa_{xy}$ is similar in all Cd-K samples; $\kappa_{xy}$ shows a peak at almost the same temperature of the peak of the phonon thermal conductivity $\kappa_{xx}^\textrm{ph}$ which is estimated by $\kappa_{xx}$ at 15\,T.
		These results indicate the presence of a dominant phonon thermal Hall $\kappa_{xy}^\textrm{ph}$ at 15\,T.
		In addition to $\kappa_{xy}^\textrm{ph}$, we find that the field dependence of $\kappa_{xy}$ at low fields turns out to be non-linear at low temperatures, concomitantly with the appearance of the field suppression of $\kappa_{xx}$, indicating the presence of a spin thermal Hall $\kappa_{xy}^\textrm{sp}$ at low fields.
		Remarkably, by assembling the $\kappa_{xx}$ dependene of $\kappa_{xy}^\textrm{sp}$ data of other kagome antiferromagnets, we find that, 
		whereas $\kappa_{xy}^\textrm{sp}$ stays a constant in the low-$\kappa_{xx}$ region, $\kappa_{xy}^\textrm{sp}$ starts to increase as $\kappa_{xx}$ does in the high-$\kappa_{xx}$ region.
		This $\kappa_{xx}$ dependence of $\kappa_{xy}^\textrm{sp}$ indicates the presence of both intrinsic and extrinsic mechanisms in the spin thermal Hall effect in kagome antiferromagnets.
		Furthermore, both $\kappa_{xy}^\textrm{ph}$ and $\kappa_{xy}^\textrm{sp}$ disappear in the antiferromagnetic ordered phase at low fields, showing that phonons alone do not exhibit the thermal Hall effect.
		A high field above $\sim$7\,T induces $\kappa_{xy}^\textrm{ph}$, concomitantly with a field-induced increase of $\kappa_{xx}$ and the specific heat, suggesting
		a coupling of the phonons to the field-induced spin excitations as the origin of $\kappa_{xy}^\textrm{ph}$.
	\end{abstract}

	\maketitle
	
	%\tableofcontents
\section{\label{sec:sec1}Introduction}

The magnetic ground state of a two-dimensional (2D) kagome structure has been attracting tremendous attention because the strong frustration effect caused by the corner-sharing network of the triangles has been expected to suppress the magnetic order even at the absolute zero temperature.
Instead of a long-range ordered state, emergence of a quantum disordered state of spins, termed as a quantum spin liquid (QSL), has been shown in the kagome Heisenberg antiferromagnet (KHA) by various numerical calculations\,\cite{YanSci2011, DepenbrockPRL2012, JiangNPhys2012, IqbalPRB2013, NishimotoNComms2013, LiaoPRL2017, HePRX2017, Jiang2019}.
A lot of QSLs have been theoretically suggested as the ground state of the KHA such as $\mathbb{Z}_2$ spin liquids\,\cite{SachdevPRB1992, WangPRB2006, DepenbrockPRL2012}, topological spin liquids\,\cite{JiangNPhys2012}, Dirac spin liquids\,\cite{IqbalPRB2013, LiaoPRL2017, HePRX2017,Jiang2019}, and chiral spin liquids\,\cite{MessioPRL2017}.
These different QSLs are characterized by different elementary excitations.
It is thus an experimental challenge to pin down the QSL realized in KHA by clarifying the elementary excitation.

Thermal-transport measurement is a powerful probe to study the elementary excitations in QSLs because it has the advantage of detecting only the itinerant excitations.
Therefore, one can avoid effects of localized excitations caused by impurities which are often inevitable in candidate materials\,\cite{FreedmanJACS2010}.
Moreover, further detail of the elementary excitation can be studied by investigating the thermal Hall effect.
It has been shown that the thermal Hall effect in an insulator is given by the Berry curvature of the elementary excitation as
\begin{eqnarray}
\label{eq_kxy}
	\kappa_{xy}=
	\frac{k_{\text{B}}T}{\hbar V}
	\sum_{\bm{k}}\sum_{n} 
	c_{2} [ g(\epsilon_{n\bm{k}}) ] \Omega_{n\bm{k}}\;,
\end{eqnarray}
where $c_{2} [ g(\epsilon_{n\bm{k}}) ] $ is a distribution function given by the elementary excitations of energy $\epsilon_{n\bm{k}}$  and $\Omega_{n\bm{k}}$ is the Berry curvature of the elementary excitations\,\cite{KatsuraPRL2010, MatsumotoPRB2014}.
Therefore, from $\kappa_{xy}$ measurements, one can study the statics of the elementary excitations (fermions or bosons) as well as the Berry curvature of the corresponding energy bands~\cite{Romhanyi2015,JHHan2019,KawanoHotta2019,Yang2020,GaoChen2020,Teng2020arXiv,FurukawaMomoi2020}.

The thermal Hall effect of spins ($\kappa_{xy}^\textrm{sp}$) has been observed in ferromagnetic insulators, which is well understood as a magnon thermal Hall effect\,\cite{OnoseSci2010, IdeuePRB2012}.
The spin thermal Hall effect has also been reported in paramagnetic states of kagome\,\cite{HirschbergerPRL2015, WatanabePNAS2016, DokiPRL2018, YamashitaJPCM2020}, spin ice~\cite{HirschbergerScience2015} and Kitaev compounds\,\cite{KasaharaPRL2018, KasaharaNat2018, HentrichPRB2019}.
In these frustrated magnets, the paramagnetic phase extends well below the temperatures determined by the interaction energy $J$, realizing a spin liquid phase in a wide temperature range $T_\textrm{N} \le T \ll J/k_\textrm{B}$.
For $\kappa_{xy}$ observed in the spin liquid phase of kagome antiferromagnets volborthite and Ca kapellasite (Ca-K), it has been shown that the Schwinger-boson mean field theory (SBMFT)\,\cite{LeePRB2015} can well reproduce both the temperature dependence and the magnitude of $\kappa_{xy}$ by tuning the two fitting parameters of the spin interaction energy $J$ and the Dzyaloshinskii-Moriya (DM) interaction $D$ (Ref.\,\cite{DokiPRL2018}).
Remarkably, the fitting results of $J$ and $D$, obtained by the SBMFT fitting to $\kappa_{xy}$ of both kagome compounds, are close to the values estimated by the temperature dependence of the magnetic susceptibility and that by the deviation of the $g$ factor, respectively. This excellent agreement suggests that the elementary spin excitations in the KHA can be well described by the bosonic spinons of SBMFT.

In addition to the spin thermal Hall effects, the thermal Hall effects of phonons ($\kappa_{xy}^\textrm{ph}$) have been reported in various compounds~\cite{Strohm2005,Sugii2017,Hirokane2019,XiaokangLi2020,Grissonnanche2020}.
The origin of the phonon thermal Hall has also been extensively studied theoretically~\cite{Sheng2006,Kagan2008,WangZhang2009,LifaZhang2010,Qin2012,MoriPRL2014,Saito2019,Zhang2019,ChenKivelsonSun2020}.
However, the understanding of the phonon thermal Hall effect has been left out in the consideration of spin thermal Hall effect, because the nature of the coupling between phonons and spin fluctuations has remained unclear.

In this Article, we report our thermal-transport measurements of a new kagome antiferromagnet Cd kapellasite (Cd-K). Previous studies\,\cite{OkumaPRB2017, OkumaNComms2019} have shown that the spin Hamiltonian of Cd-K is well approximated to a KHA with the spin interaction energy of $J/k_{\text{B}}\sim 45$ K.
The frustration effect of the kagome structure suppresses the ordering temperature ($ T_\textrm{N} \sim 4$\,K) well below $J/k_{\text{B}}$, realizing a spin liquid phase in a wide temperature range.
\textcolor{black}{
	We find a large spin contribution in $\kappa_{xx}$ which can be strongly suppressed by applying a magnetic field. This field suppression effect on $\kappa_{xx}^\textrm{sp}$ allows us to identify both $\kappa_{xy}^\textrm{sp}$ and $\kappa_{xy}^\textrm{ph}$ in Cd-K.
	Most remarkably, we find the $\kappa_{xx}$ dependence of $\kappa_{xy}^\textrm{sp}$ indicates the presence of both intrinsic and extrinsic mechanisms depending on the strength of the impurity scatterings for the spin thermal Hall effect.
	Furthermore, we find that both $\kappa_{xy}^\textrm{sp}$ and $\kappa_{xy}^\textrm{ph}$ disappear in the AFM phase at low fields.
	Applying a high field in the AFM phase induces $\kappa_{xy}^\textrm{ph}$, concomitantly with the appearance of additional excitations probed by the specific heat and $\kappa_{xx}$.
	We conclude that Cd-K is a prominent frustrated magnet in which the spin liquid state shows thermal Hall effects of both spins and phonons. 
	The dual nature of the thermal conductivities prompts us to speculate that a spin-phonon coupling gives rise to both $\kappa_{xy}^\textrm{sp}$ and $\kappa_{xy}^\textrm{ph}$ in this compound.
}

\section{\label{sec2}Materials and Methods}

\begin{figure}[bt]
	\includegraphics[width=8.6cm]{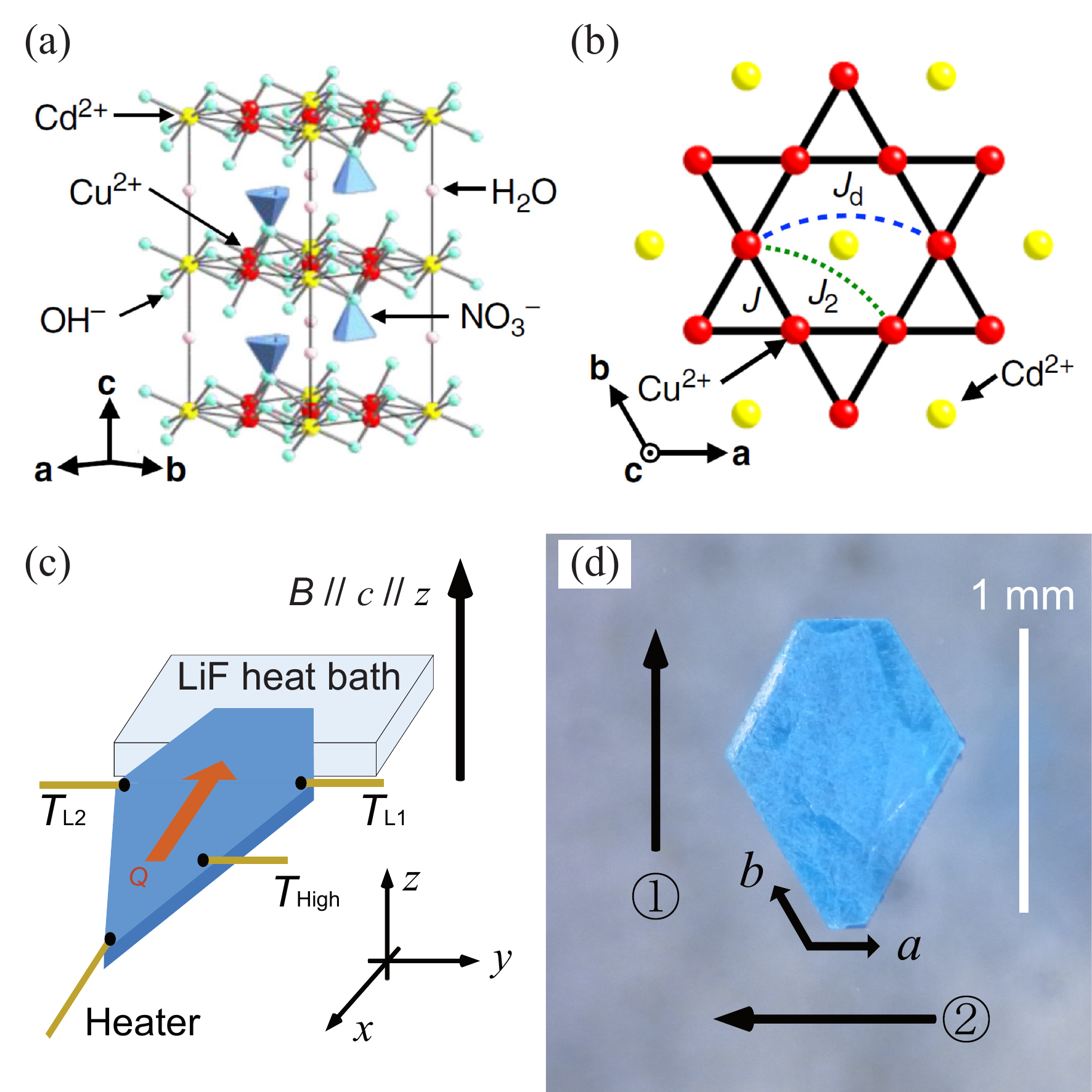}
	\caption{\label{fig:sample} (a) Crystal structure and (b) top-view of a kagome layer of Cd-K\,\cite{OkumaNComms2019}. The magnetic interactions between nearest-neighbor, next-nearest-neighbor, and diagonal Cu$^{2+}$ spins are denoted by $J$ (solid black line),  $J_2$ (dotted line) and $J_\textrm{d}$ (dashed line), respectively. (c) Schematic illustration of $\kappa_{xx}$ and $\kappa_{xy}$ measurements. A heater and three thermometers ($T_\textrm{High}$, $T_\textrm{L1}$, $T_\textrm{L2}$) were attached to the sample fixed on the LiF heat bath. A heat current $Q$ was applied within the kagome layer and a magnetic field $B$ was applied along the $c$-axis. (d) A typical crystal of Cd-K. The direction of the heat current for Sample 1, 2, and 3-1 (3-2) is shown by the arrow 1 (2).}
\end{figure}

Cd-Kapellasite CdCu$_3$(OH)$_6$(NO$_3$)$_{2}$\,$\cdot$\,H$_2$O is a trigonal compound with space group $P\overline{3}m1$ and lattice constants $a=6.5449$ \AA , $c=7.0328$ \AA\,\cite{OkumaPRB2017}, in which the magnetic Cu$^{2+}$ ions form an undistorted kagome lattice (Fig.\,\ref{fig:sample} (a)). 
Cd-kapellasite is isostructural to Zn-kapellasite ZnCu$_3$(OH)$_6$Cl$_2$ which is a polymorph of herbertsmithite~\cite{Shores2005}.
In herbertsmithite, the Zn ions located between the kagome layer and the site mixings between the Zn and Cu ions~\cite{FreedmanJACS2010} allow an inter-layer coupling between the kagome layers. In contrast, in Cd-K, the nonmagnetic Cd ions are located at the center of the hexagon of the kagome lattice and there is no site mixings in Cd-K because of the larger ionic radii of Cd$^{2+}$ (0.95 \AA) than that of Cu$^{2+}$ (0.73 \AA)\,\cite{ShannonActaA1976}, realizing a more ideal KHA in Cd-K.

Three magnetic interactions in Cd-K are suggested by the fitting of the magnetic susceptibility as $J/k_{\text{B}}=45.44$ K, $J_{2}/J=-0.1$, $J_{d}/J=0.18$, where $J$ is the nearest-neighbor interaction, $J_2$ the next-nearest-neighbor interaction, and $J_\textrm{d}$ the diagonal interaction via the non-magnetic Cd ion (see Fig.\,\ref{fig:sample} (b)). The development of a short-range antiferromagnetic correlation has been shown by the decrease of the magnetic susceptibility below 30 K\,\cite{OkumaPRB2017}. The $g$ factors are estimated as $g_{a}=2.28$, $g_{c}=2.37$\,\cite{OkumaNComms2019}. The lack of the inversion symmetry allows both the in-plane and out-of-plane DM interactions, which has been suggested to cause a negative vector chiral order below the Neel temperature of $T_{\text{N}}\sim 4$ K\,\cite{OkumaPRB2017}.

The thermal-transport measurements were performed by the steady-state method as described in Refs.\,\onlinecite{WatanabePNAS2016, DokiPRL2018, YamashitaJPCM2020}. One heater and three thermometers were attached to the sample, and then the temperature difference $\Delta T_x$ ($\Delta T_{x}=T_\textrm{High}-T_\textrm{L1}$) and $\Delta T_y$ ($\Delta T_{y}=T_\textrm{L1}-T_\textrm{L2}$) were measured by applying the heat current $Q$ in the kagome plane (Fig.\,\ref{fig:sample} (c)). 
The longitudinal thermal conductivity $\kappa_{xx}$ and the thermal Hall conductivity $\kappa_{xy}$ is derived by
\begin{eqnarray}
\left(
\begin{array}{c}
Q/wt\\
0
\end{array}\right)=
\left(
\begin{array}{cc}
\kappa_{xx} & \kappa_{xy}\\
-\kappa_{xy} & \kappa_{xx}
\end{array}\right)
\left(
\begin{array}{c}
\Delta T_{x}/L\\
\Delta T_{y}^\textrm{Asym}/w'
\end{array}\right)\;,
\label{eq:kxxkxy}\end{eqnarray}
where $t$ is the thickness of the sample, $L$ is the length between $T_\textrm{High}$ and $T_\textrm{L1}$, $w$ is the averaged sample width between $T_\textrm{L1}$ and $T_\textrm{L2}$, $w'$ is the length between $T_\textrm{L1}$ and $T_\textrm{L2}$, and $\Delta T_{y}^\textrm{Asym}$ is the antisymmetrized $\Delta T_y$ with respect to the field direction as $\Delta T_{y}^\textrm{Asym}(B) =(\Delta T_{y}(+B) - \Delta T_{y}(-B))/2$.

We measured $\kappa_{xx}$ and $\kappa_{xy}$ of three Cd-K samples (Sample 1, 2, and 3) by using a variable temperature insert (VTI) (2--60\,K, 0--15\,T). Measurements of Sample 2 were also done in a dilution refrigerator (DR) (0.1--4\,K, 0--14\,T). The magnetic field was applied along the $c$ axis of the sample. 
A typical sample is shown in Fig.\,\ref{fig:sample}(d). 
Because of the non-rectangular shape of the sample, there is an ambiguity up to about 40\% in estimating the absolute value of $\kappa_{xx}$ and $\kappa_{xy}$ (see Supplemental Material (SM)~\cite{SM} for more details).
A heat current $Q$ was applied along the direction 1 ($\perp a$-axis, see Fig.\,\ref{fig:sample}(d)) in Samples 1, 2, and the first run of Sample 3 (denoted as Sample 3-1). In the second run of Sample 3 (Sample 3-2), the direction of $Q$ was changed to the direction 2 ($\parallel a$-axis, see Fig.\,\ref{fig:sample} (d)). 
	For each $\kappa_{xx}$ and $\kappa_{xy}$ measurements, we confirmed the linear $Q$ dependence of both $\Delta T_x$ and  $\Delta T_{y}^\textrm{Asym}$ (see Fig.\,S1 in SM~\cite{SM}). We also checked the temperature stability during the measurements is good enough to resolve $\Delta T_{y}^\textrm{Asym}$ (Fig.\,S2 in SM~\cite{SM}).

The specific heat measurements were performed for two sets of multiple single crystals by a thermal relaxation method by using a Physical Property Measurement System (PPMS, Quantum Design) and a DR.
The PPMS measurements (2--10\,K, 0--10\,T) were performed for the same set of the single crystals used in Ref.\,\cite{OkumaPRB2017}.
The DR measurements (0.1--2\,K, 0--14\,T) were performed for another set of single crystals.
The magnetic field was applied along the $c$ axis of the samples in all the measurements. 

%%%%%%%%%%%%%%%%%%%%%%%%%%%%%%%%%%%%%%%%%
%%			RESULTS
%%
\section{\label{sec3}Results}

\subsection{\label{Res_kxx}Longitudinal Thermal Conductivity}

\begin{figure*}
	\includegraphics[width=0.9\linewidth]{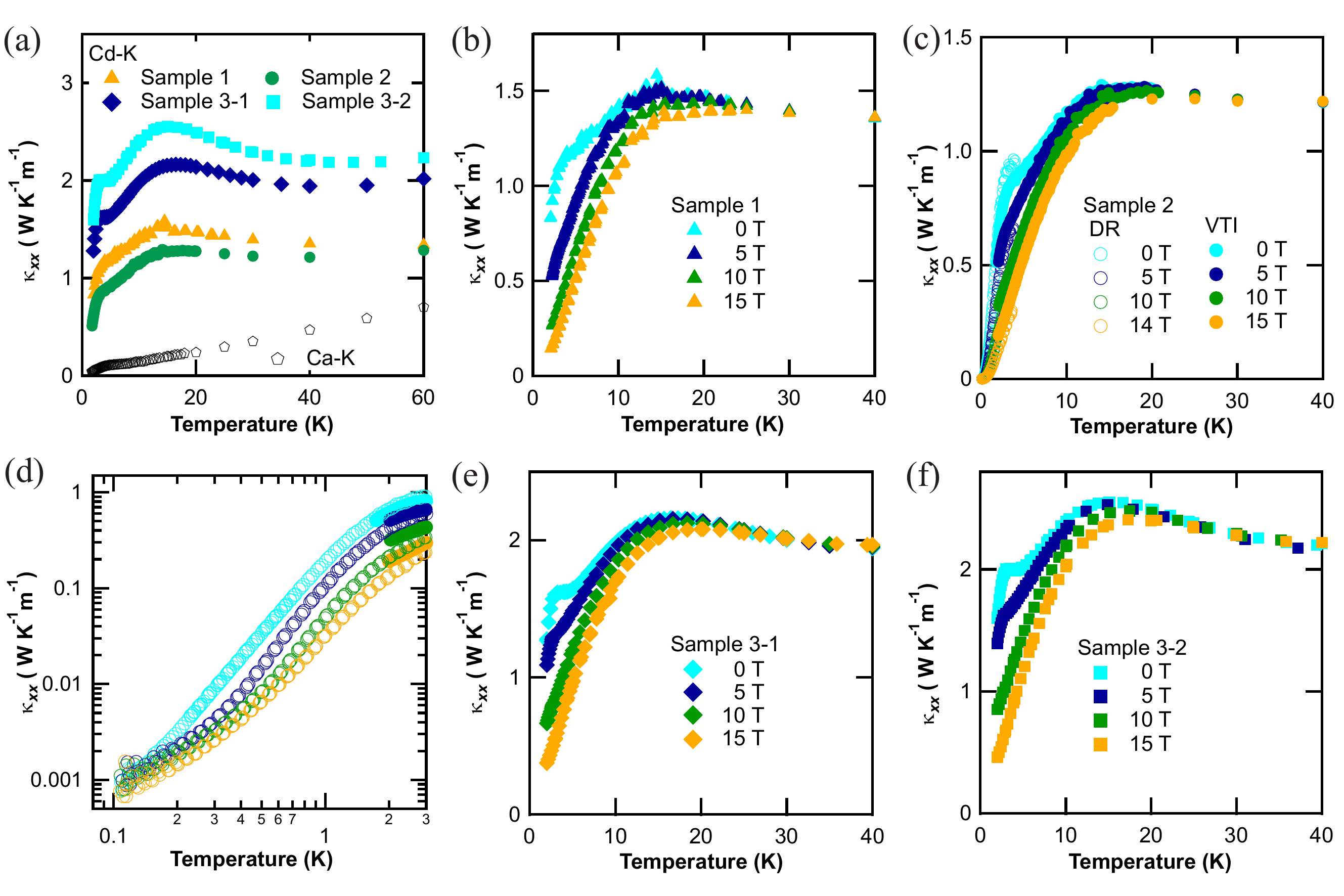}% Here is how to import EPS art
	\caption{\label{fig:kxx_vs_T} (a) Temperature dependence of the longitudinal thermal conductivity ($\kappa_{xx}$) of all samples of Cd-Kapellasite (Cd-K) and that of Ca-Kapellasite (Ca-K) at 0 T. The longitudinal thermal conductivity of Ca-K is taken from Ref. \cite{DokiPRL2018}.
		(b to f) The data of each sample under magnetic fields.
		In Sample 2, $\kappa_{xx}$ was measured down to 0.1 K.
		An enlarged view of low-temperature region (0.1--3 K) of (c) is shown in (d). The filled and open circles in (c) and (d) show $\kappa_{xx}$ measured by a variable temperature insert (VTI) (2 K$<T$) and a dilution refrigerator (DR) ($0.1<T<4$ K), respectively. 
		The slight difference between the two data might be caused by a thermal cycle effect and/or different setups between the VTI and the DR measurements. 
	}
\end{figure*}

Figure \ref{fig:kxx_vs_T} (a) shows the temperature dependence of $\kappa_{xx}$ of all Cd-K samples at zero magnetic field. For reference, $\kappa_{xx}$ of Ca-K\,\cite{DokiPRL2018} is also shown. As shown in Fig.\,\ref{fig:kxx_vs_T} (a), $\kappa_{xx}$ of all Cd-K samples is about one order of magnitude larger than that of Ca-K. Although the magnitude of $\kappa_{xx}$ in different Cd-K samples are different in factor of $\sim 2$, $\kappa_{xx}$ of all Cd-K samples show a similar temperature dependence. The temperature dependence of $\kappa_{xx}$ shows a shoulder-like enhancement around 15 K, which is followed by a hump near $T_{\text{N}}$ and a rapid decrease for $T<T_{\text{N}}$.

The temperature dependence of $\kappa_{xx}$ at different magnetic fields is shown in Figs.\,\ref{fig:kxx_vs_T} (b)--(f). In all Cd-K samples, a decrease of $\kappa_{xx}$ by applying the magnetic field was observed below $\sim$25\,K. This field suppression effect is larger for a sample with a large $\kappa_{xx}$. 
In Sample 2, additional lower temperature measurements were performed by using a DR (open circles in Fig.\,\ref{fig:kxx_vs_T}(c) and (d)). As shown in Fig.\,\ref{fig:kxx_vs_T}(d), a large field suppression effect was observed at $\sim$1\,K.

\begin{figure*}
	\includegraphics[width=0.9\linewidth]{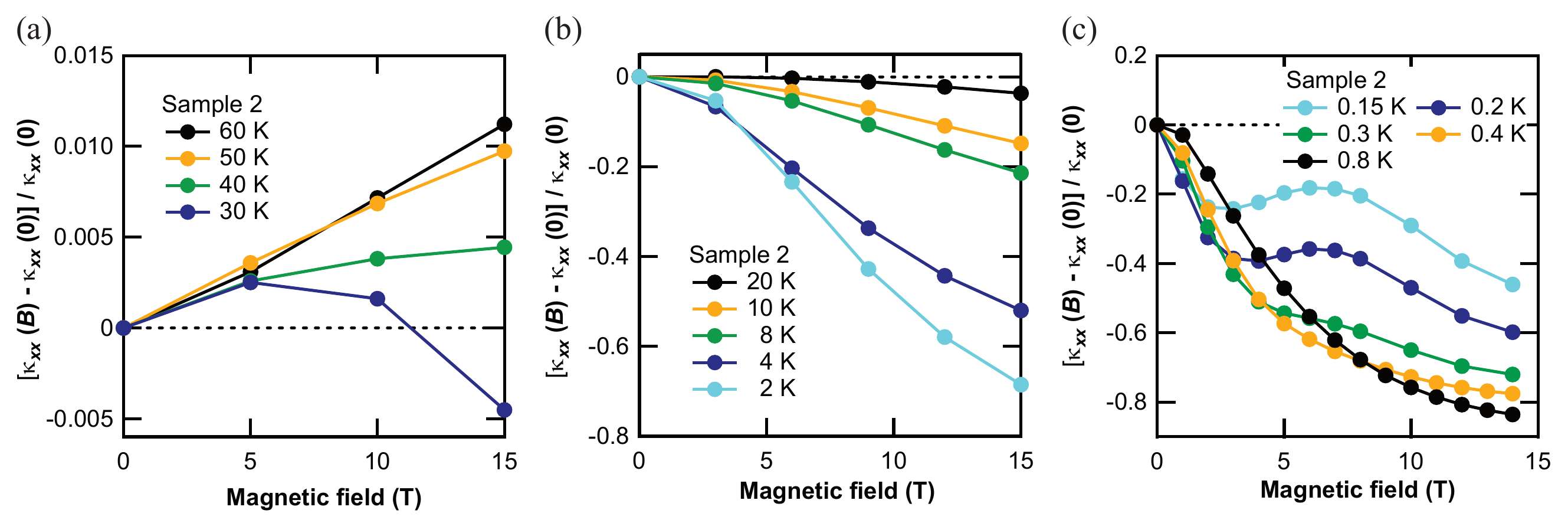}
	\caption{\label{fig:knrm_vs_B} Magnetic field dependence of the longitudinal thermal conductivity normalized by the zero field value $([\kappa_{xx}(B) - \kappa_{xx}(0)]/\kappa_{xx}(0))$ of Sample 2 above 30 K (a), for 2--20 K (b), and below 2 K (c).}
\end{figure*}

Figure \ref{fig:knrm_vs_B} shows the magnetic field dependence of $\kappa_{xx}$ of Sample 2. The vertical axis is normalized by the zero-field value as $[\kappa_{xx}(B)-\kappa_{xx}(0)]/\kappa_{xx}(0)$. The field dependence of $\kappa_{xx}$ of other samples were essentially the same. Above 40 K, $\kappa_{xx}$ increased linearly by applying the magnetic field (Fig.\,\ref{fig:knrm_vs_B} (a)). On the other hand, below $\sim$25\,K, the suppression of $\kappa_{xx}$ by the magnetic field was observed (Fig.\,\ref{fig:knrm_vs_B} (b)). The field suppression effect became larger at lower temperatures and reached the maximum reduction of $\sim$80\% by 15\,T at 1\,K. Below 0.3,K, a new peak was observed in the field dependence of $\kappa_{xx}$ at 6--7\,T (Fig.\,\ref{fig:knrm_vs_B} (c)).

\subsection{\label{Res_kxy}Thermal Hall Conductivity}

\begin{figure*}[!hbt]
	\includegraphics[width=0.9\linewidth]{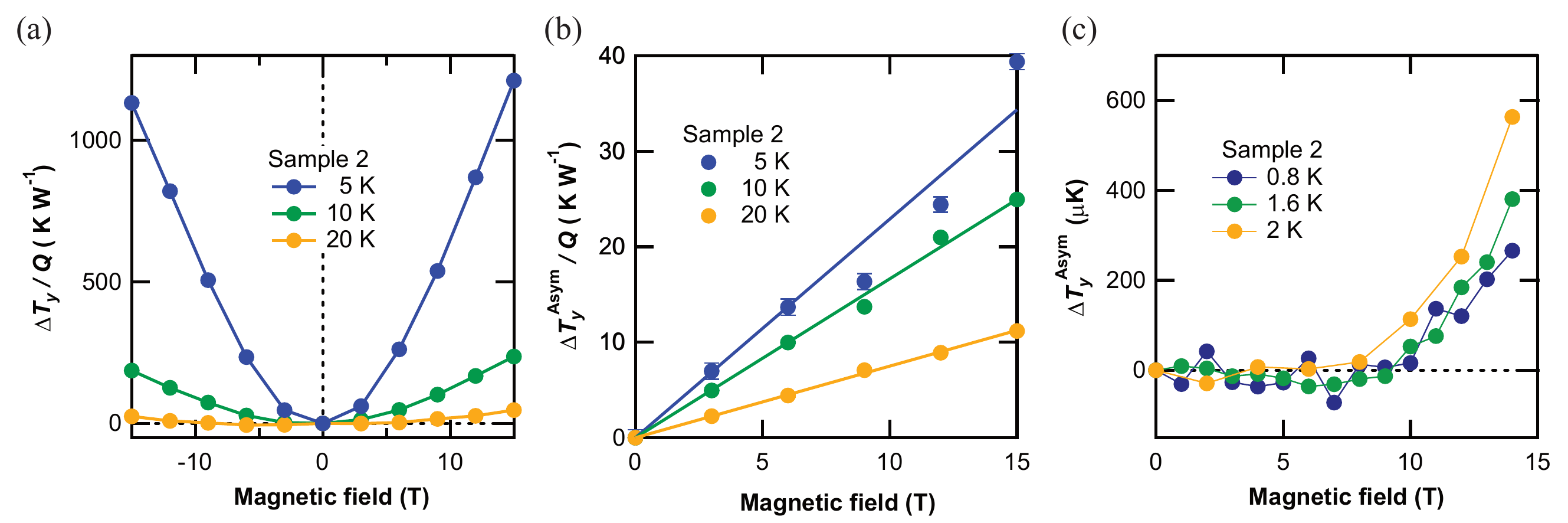}
	\caption{\label{fig:dTy_vs_B} 
		(a) Magnetic field dependence of the transverse temperature difference divided by the heat current $(\Delta T_{y}/Q)$. The zero-field value of $\Delta T_{y}/Q$, which is caused by the misalignment effect, is subtracted for clarity.
		(b) Magnetic field dependence of the asymmetrized $\Delta T_{y}/Q$ of (a) with respect to the field direction. See text for details. The solid lines represent a linear fitting to $\Delta T_{y}^\textrm{Asym}/Q$ for each temperature. 
		(c) Magnetic field dependence of $\Delta T_{y}^\textrm{Asym}$ below 2 K. Error bars represent the standard error of the data, which are smaller than the symbol size except for the 5\,K data in (b).
	}
\end{figure*}

Figure \ref{fig:dTy_vs_B}(a) shows the magnetic field dependence of $\Delta T_{y}/Q$ of Sample 2 in the spin liquid phase. As shown in Fig.\,\ref{fig:dTy_vs_B}(a), the field dependence of $\Delta T_{y}/Q$ is dominated by the symmetric longitudinal component caused by the misalignment effect. To extract the asymmetric thermal Hall effect, the field dependence of $\Delta T_{y}/Q$ is antisymmetrized with respect to the field direction as $\Delta T_{y}^\textrm{Asym}(B) =(\Delta T_{y}(+B) - \Delta T_{y}(-B))/2$. 
The field dependence of  $\Delta T_{y}^\textrm{Asym}(B)$ of Sample 2 is shown in Fig.\,\ref{fig:dTy_vs_B}(b).
As shown in Fig.\,\ref{fig:dTy_vs_B}(b), $\Delta T_{y}^\textrm{Asym}/Q$ shows a linear magnetic field dependence at high temperatures.

\begin{figure*}[!hbt]
	\includegraphics[width=0.9\linewidth]{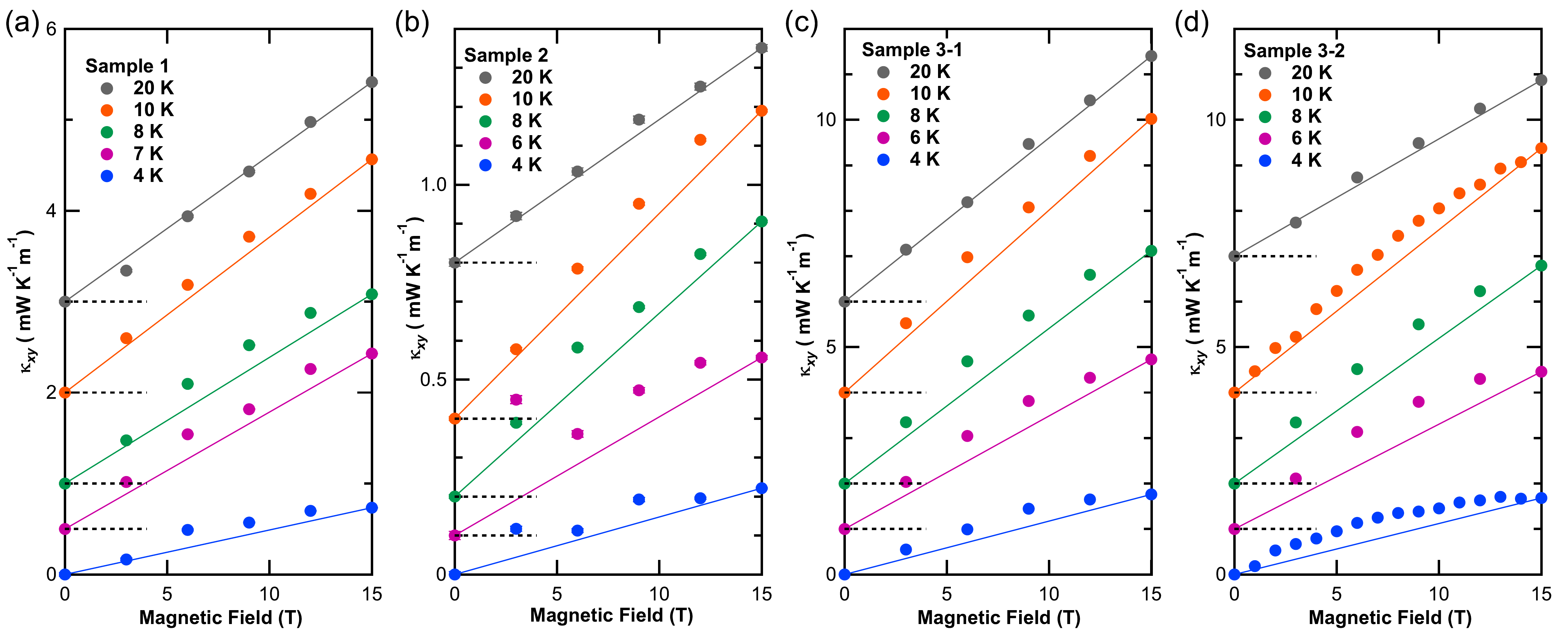}
	\caption{\label{fig:k_xy_vs_B_all_samples} 
		The field dependence of $\kappa_{xy}$ of all Cd-K samples for 4--20\,K. The data above 4\,K is shifted for clarity.
		The offsets for the shifted data are indicated by the dashed lines.
		\textcolor{black}{
			The solid lines are drawn to show a deviation from the linear increase of $\kappa_{xy}$ to that at 15\,T.
		}
		Error bars estimated by the standard error are smaller than the symbol size for all the measurements.
	}
\end{figure*}

The field dependence of $\kappa_{xy}$ is determined by $\Delta T_{y}^\textrm{Asym}$ in accordance with Eq.\,\ref{eq:kxxkxy}, and is plotted in Figs.\,\ref{fig:k_xy_vs_B_all_samples}.
As shown in Figs.\,\ref{fig:k_xy_vs_B_all_samples}, the linear field dependence of $\kappa_{xy}$ (solid lines in Figs.\,\ref{fig:k_xy_vs_B_all_samples}) was observed at 20\,K, which became non-linear at lower temperatures.

\begin{figure}[!hbt]
	\includegraphics[width=8.6cm]{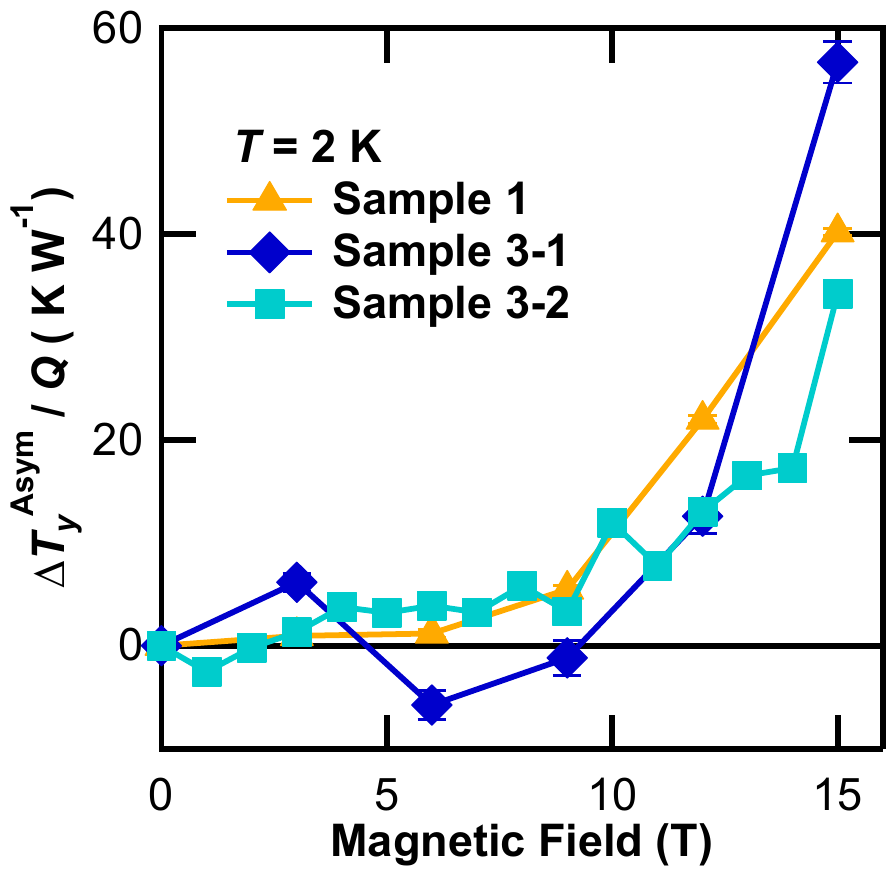}
	\caption{\label{fig:dTy_vs_B_allS} Magnetic field dependence of the asymmetrized transverse temperature difference divided by the heat current $(\Delta T_{y}^\textrm{Asym}/Q)$ at 2\,K.  Error bars represent the standard error of the measurements.
	}
\end{figure}

This thermal Hall signal disappears in the AFM phase at low fields.
Figure\,\ref{fig:dTy_vs_B}(c) shows the field dependence of $\Delta T_{y}^\textrm{Asym}$ of Sample 2 measured in a dilution refrigerator. 
As shown in Fig.\,\ref{fig:dTy_vs_B}(c), the thermal Hall effect was absent at low fields, which is followed by an increase above $\sim 7$\,T.
A similar non-linear field dependence is confirmed in all Cd-K samples done at the lowest temperature of the VTI measurement (2\,K) as shown in Fig.\,\ref{fig:dTy_vs_B_allS}.

\begin{figure*}[!hbt]
	\includegraphics[width=0.9\linewidth]{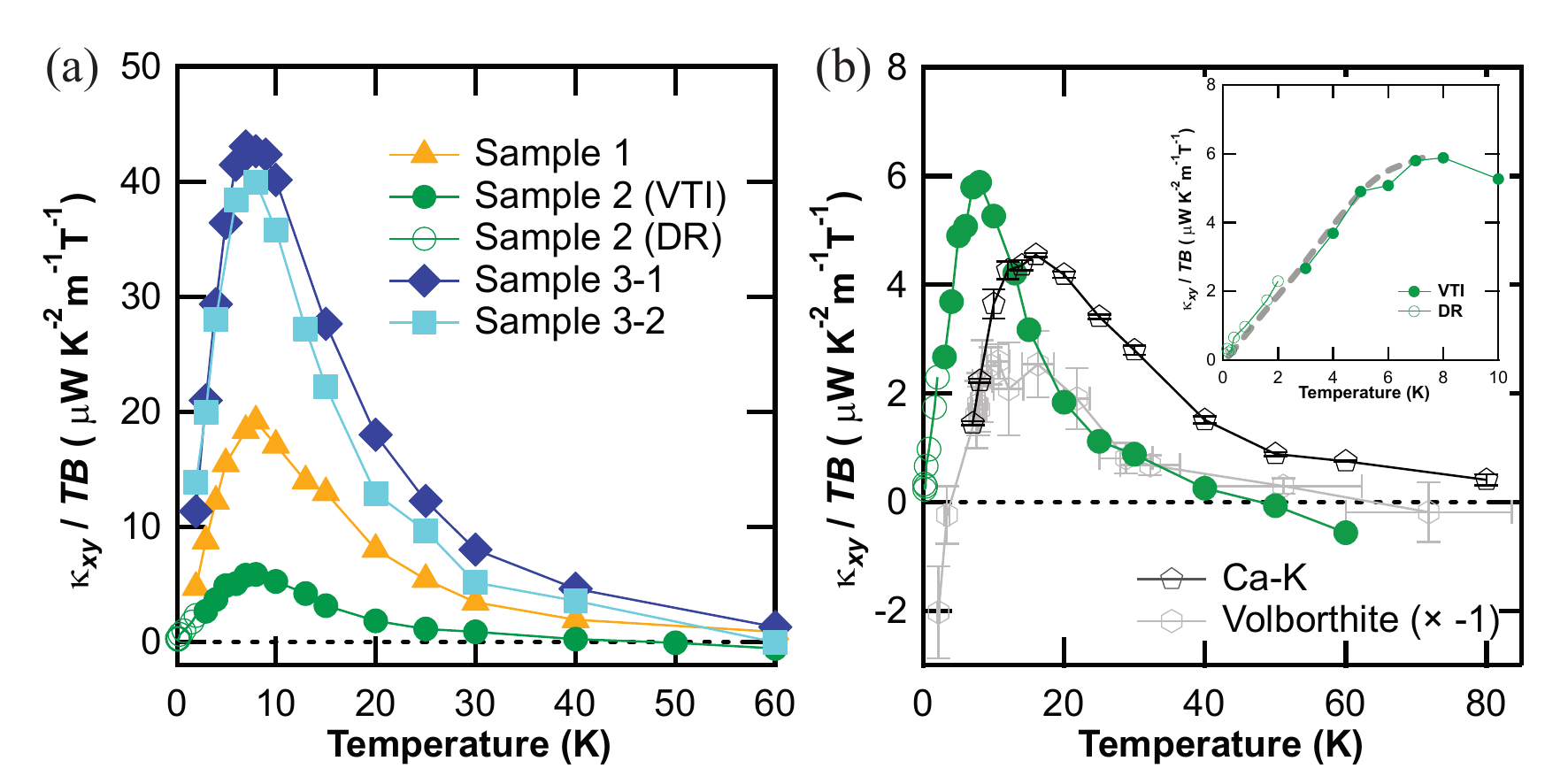}
	\caption{\label{fig:kxy_vs_T} Temperature dependence of $\kappa_{xy}/TB$. (a) Comparison of $\kappa_{xy}/TB$ of three Cd-K samples. 
	The filled (open) symbols represent $\kappa_{xy}/TB$ at 15\,T in the VTI measurements (14\,T in the DR measurements). 
	(b) Comparison of $\kappa_{xy}/TB$ of Cd-K Sample 2 (green circles), Ca-K\,\cite{DokiPRL2018} (open gray pentagon) and volborthite\,\cite{WatanabePNAS2016} (open gray hexagon). For clarity, $\kappa_{xy}/TB$ of volborthite is multiplied by $-1$. 
	The inset shows an enlarged view of the low temperature data of $\kappa_{xy}/TB$ of Sample 2.
	The dashed line shows a guide to the eye.
	}
\end{figure*}

Figure \ref{fig:kxy_vs_T} (a) shows the temperature dependence of $\kappa_{xy}/TB$ of all Cd-K samples at 15 (14)\,T for the VTI (DR) measurements. As shown in Fig.\,\ref{fig:kxy_vs_T}(a),  $\kappa_{xy}/TB$ of all Cd-K samples shows a similar temperature dependence with a peak around 8 K. In Sample 2, $\kappa_{xy}/TB$ obtained in the ordered phase at 14\,T done by the DR measurements is also shown by open symbols, which seems to be smoothly connected to the VTI data shown by filled symbols (see the inset of Fig.\,\ref{fig:kxy_vs_T}(b)).
This temperature dependence is also similar to that of Ca-K\,\cite{DokiPRL2018} and volborthite \,\cite{WatanabePNAS2016} (Fig.\,\ref{fig:kxy_vs_T}(b)). On the other hand, as shown in Fig.\,\ref{fig:kxy_vs_T}(b), the peak temperature of $\kappa_{xy}/TB$ is clearly shifted to a lower temperature in Cd-K.

\subsection{\label{Res_C}Specific Heat}

\begin{figure*}[!hbt]
	\includegraphics[width=0.9\linewidth]{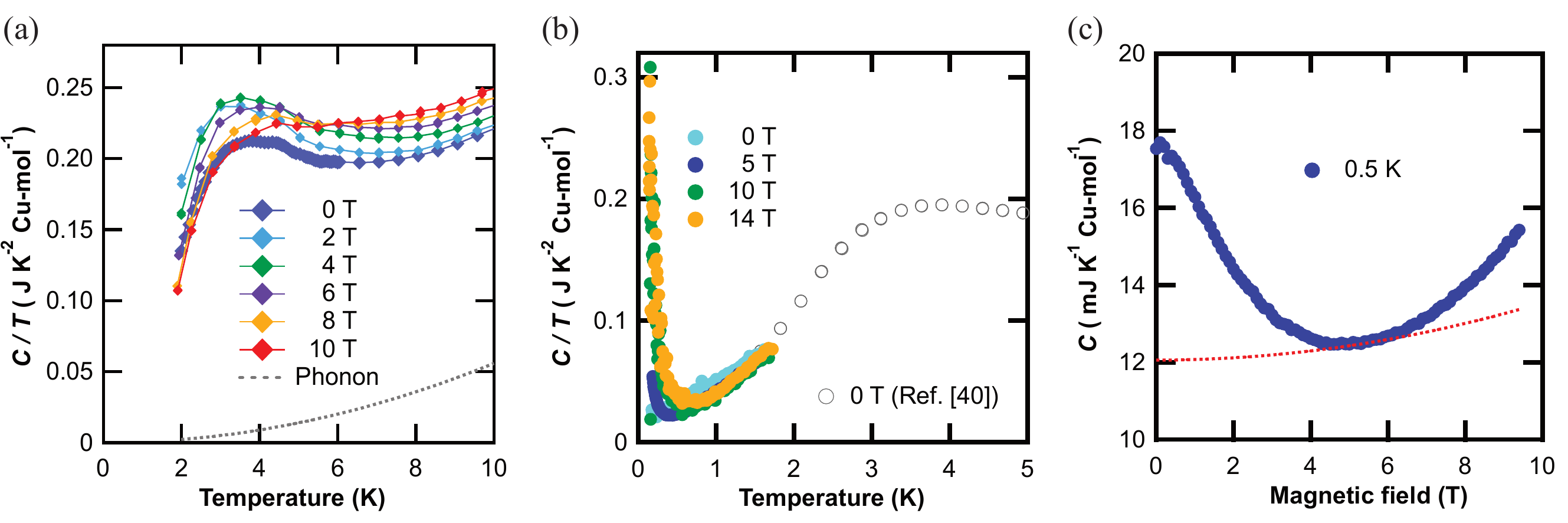}
	\caption{\label{fig:Specifit_Heat} 
		(a) Temperature dependence of the specific heat divided by the temperature ($C/T$) measured in PPMS. The measurements were done for the same set of the multiple single crystals used in Ref\,\cite{OkumaPRB2017}. The peak around 4\,K corresponds to the AFM transition.
		The dashed line shows the phonon contribution $C_\textrm{ph}/T$ estimated by a fitting for the data at high temperatures (see SM~\cite{SM} for details).
		(b) Temperature dependence of $C/T$ of another set of multiple single crystals measured in DR. The zero-field data from the previous report~\cite{OkumaPRB2017} is also shown by open circles.
		(b) Magnetic field dependence of $C$ at 0.5 K. 
		The red dotted line shows an estimation of the magnetic field dependence of the nuclear Schottky specific heat ($C_\textrm{Ncl}$) at 0.5 K. The data of $C_\textrm{Ncl}$ is shifted to compare the amount of the field increase.
	}
\end{figure*}

	Figure \ref{fig:Specifit_Heat}(a) shows the temperature dependence of the specific heat divided by the temperature  ($C/T$) measured in PPMS.
	As shown in Fig.\,\ref{fig:Specifit_Heat}(a), although the peak in $C/T$ by the AFM transition becomes broader at higher fields, the peak temperature does not depend on the magnetic field up to 5\,T, which is followed by a slight increase of $T_\textrm{N}$.

	Because of the absence of a non-magnetic compound isostructural to Cd-K, we estimate the phonon specific heat $C_\textrm{ph}/T$ by fitting the data at high temperatures (see SM~\cite{SM} for details).
	As shown by the dashed line in Fig.\,\ref{fig:Specifit_Heat}(a), $C_\textrm{ph}/T$ is considerably smaller than $C/T$ below 10\,K, showing that the specific heat is dominated by the magnetic contribution.

Figure \ref{fig:Specifit_Heat}(b) shows the temperature dependence of $C/T$ of another set of multiple single crystals measured in DR. The zero-field data shows a good agreement with the previous data\,\cite{OkumaPRB2017} shown as open circles in Fig.\,\ref{fig:Specifit_Heat}(b). Below 0.5\,K, $C/T$ under magnetic fields increases rapidly as lowering temperature owing to the nuclear Schottky anomaly ($C_\textrm{Ncl}$). The magnetic field dependence of the specific heat at 0.5 K is shown in Fig.\,\ref{fig:Specifit_Heat}(c). After the specific heat was decreased by applying the magnetic field, the specific heat was increased by applying the magnetic field above 7\,T. We note that this field increase of $C$ above 7\,T is much larger than that expected by $C_\textrm{Ncl}$ (dotted line in Fig.\,\ref{fig:Specifit_Heat}(c)) which is estimated as $\sim 2$ mJ K$^{-1}$ mol$^{-1}$ at 0.5 K and 10 T from the fit of $C_\textrm{Ncl}\propto H^{2}/T^{2}$ for the data shown in Fig.\,\ref{fig:Specifit_Heat}(b).

\newpage

\subsection{$B$--$T$ phase diagram of Cd-K} \label{sec:PhaseDiag}

\begin{figure}[!hbt]
	\includegraphics[width=0.7\linewidth]{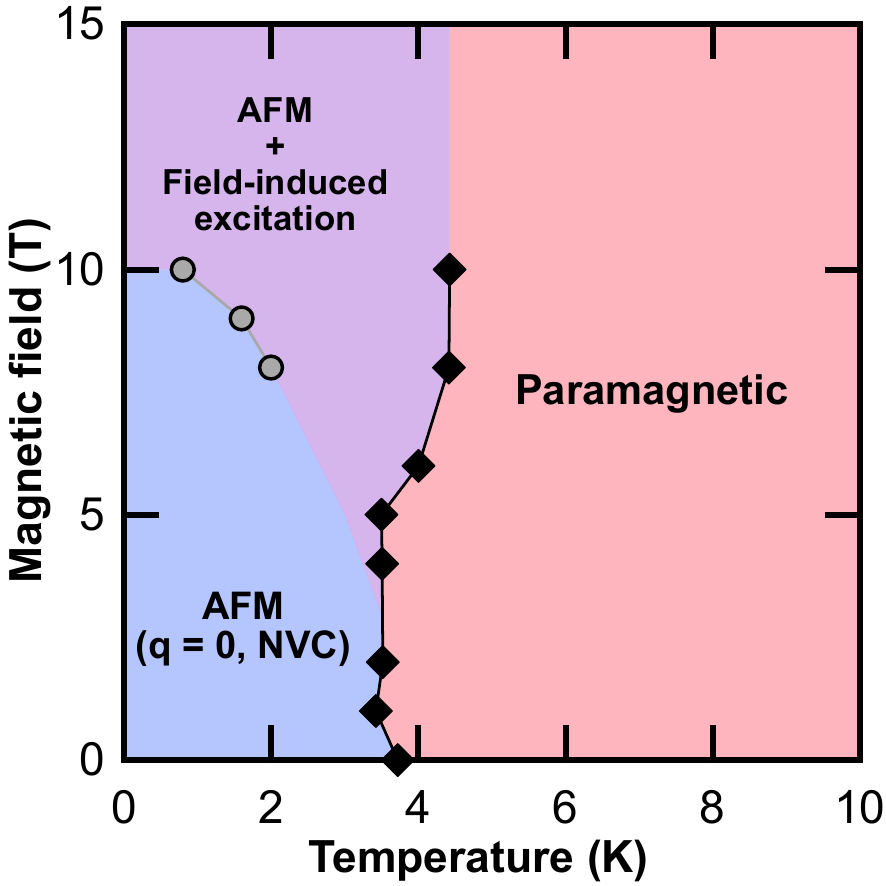}
	\caption{	\label{fig:PhaseDiag}
		$B$--$T$ phase diagram of Cd-K.
		The boundary of the antiferromagnetic (AFM) phase is determined by the peak temperature of $C/T$ (black diamonds) shown in Fig.\,\ref{fig:Specifit_Heat}.
		The threshold fields where the onset of the finite $\Delta T_{y}^\textrm{Asym}$ is observed in the field dependence (Fig.\,\ref{fig:dTy_vs_B}(c)) are shown by grey circles.
	}
\end{figure}

	From the specific heat measurements at different magnetic fields (Fig.\,\ref{fig:Specifit_Heat}(a)) and the field dependence of $\Delta T_{y}^\textrm{Asym}$ (Fig.\,\ref{fig:dTy_vs_B}(c)), we determined the $B$--$T$ phase diagram of Cd-K (Fig.\,\ref{fig:PhaseDiag}). As shown in Fig.\,\ref{fig:PhaseDiag}, $T_\textrm{N}$ determined by the peak temperature of $C/T$ slightly increases above 5\,T. A similar increase of $T_{\text{N}}$ has been observed in kapellasite~\cite{Kermarrec2014} and Ca-K~\cite{Ihara2020}. The threshold field of $\Delta T_{y}^\textrm{Asym}$ (grey circles) seems to gradually decrease to zero as $T$ increases to $T_{\text{N}}$.

\newpage

%%%%%%%%%%%%%%%%%%%%%%%%%%%%%%%%%%%%%%%
%%
%%	DISCUSSION
%%

\section{Discussion} \label{sec:Discussion}

\subsection{Longitudinal Thermal Conductivity} \label{Dis:kxx}

First, we discuss the sample dependence of $\kappa_{xx}$ (Fig.\,\ref{fig:kxx_vs_T} (a)) in terms of the sample quality. The longitudinal thermal conductivity of an insulator is given by the sum of the contribution of the phonons $\kappa_{xx}^\textrm{ph}$ and that of the spins $\kappa_{xx}^\textrm{sp}$ \cite{FNdecomposing}. 
Considering $J/k_{\text{B}}\sim 45$ K, it can be expected that $\kappa_{xx}$ above 45 K is almost given by $\kappa_{xx}^\textrm{ph}$, which is consistent with the field dependence of $\kappa_{xx}$. It is known that $\kappa_{xx}^\textrm{ph}$ increases in the magnetic field because the spin-phonon scatterings are reduced under magnetic field by suppressing spin fluctuations\,\cite{Berman1976}. In fact, as shown in Fig.\,\ref{fig:knrm_vs_B} (a), the increase of $\kappa_{xx}$ by applying the magnetic field is observed above 40 K, showing a dominant $\kappa_{xx}^\textrm{ph}$ in  $\kappa_{xx}$ at high temperatures.

The phonon thermal conductivity $\kappa_{xx}^\textrm{ph}$ is given by a product of the specific heat $C_\textrm{ph}$, the mean free path $\ell_\textrm{ph}$, and the velocity $v_\textrm{ph}$ of phonons, as $\kappa_{xx}^\textrm{ph}=(1/3)C_\textrm{ph}l_\textrm{ph}v_\textrm{ph}$. Since $C_\textrm{ph}$ and $v_\textrm{ph}$ are common in all Cd-K samples, the difference in the magnitude of $\kappa_{xx}$ shown in Fig.\,\ref{fig:kxx_vs_T} (a) reflects the difference in $\ell_\textrm{ph}$ of each sample. Therefore, a sample with a larger $\kappa_{xx}$ is a better crystal with less impurities. Also, the larger $\kappa_{xx}$ of Cd-K than that of Ca-K shows that $\ell_\textrm{ph}$ of Cd-K is much longer than that of Ca-K because $C_\textrm{ph}$ and $v_\textrm{ph}$ of Cd-K are similar to those of the isostructural  Ca-K.
This longer $\ell_\textrm{ph}$ of Cd-K than that of Ca-K~\cite{DokiPRL2018} indicates that Cd-K has a more ideal kagome structure without the randomness of ions or the lattice defects found  in Ca-K\,\cite{YoshidaJPSJ2017}. 
We note that the difference of $\kappa_{xx}$ of Sample 3-1 and that of Sample 3-2 might be caused by the ambiguity in estimating the sample size (up to 10\%) owing to the irregular shape of the sample (see Fig.\,\ref{fig:sample} (d)).

Next, we discuss the field suppression effect on $\kappa_{xx}$ observed below $\sim$25\,K (Fig.\,\ref{fig:knrm_vs_B} (b)). One of the field-suppression mechanisms of $\kappa_{xx}^\textrm{ph}$, which normally increases in the magnetic field, is a resonance scattering of phonons being absorbed by impurity free spins\,\cite{Berman1976}. This resonant scattering is most effective when the spin Zeeman gap ($g\mu_{B}H$) coincides with the phonon peak ($\sim 4k_{\text{B}}T$) given by the Debye distribution, where $\mu_\textrm{B}$ is the Bohr magneton.
Therefore, this resonance scattering produces a suppression peak of $\kappa_{xx}$ at 5.4\,T for 2\,K as observed in volborthite~\cite{WatanabePNAS2016}.
However, as shown in Fig.\,\ref{fig:knrm_vs_B}(b), $\kappa_{xx}$ at 2 K decreases monotonically with increasing magnetic field up to 15 T without the expected suppression peak. Therefore, the field suppression effect of $\kappa_{xx}$ cannot be explained by the resonance scattering effect on $\kappa_{xx}^\textrm{ph}$. Therefore, the field suppression effect is caused by the decrease of $\kappa_{xx}^\textrm{sp}$ under magnetic fields. A similar field suppression effect on $\kappa_{xx}^\textrm{sp}$ has also been observed in the spin liquid state of the one-dimensional (1D) spin-chain compound~\cite{Sologubenko2007}, volborthite \,\cite{WatanabePNAS2016, YamashitaJPCM2020} and Ca-K\,\cite{DokiPRL2018}. In volborthite, a field suppression effect up to $\sim$30\% at 15\,T was observed\,\cite{YamashitaJPCM2020} together with the resonance scattering effect\,\cite{WatanabePNAS2016}. Compared to the field suppression effects in volborthite and Ca-K, the field suppression of $\kappa_{xx}$ in Cd-K is much larger ($\sim$80\% at $\sim$1\,K), showing a dominant contribution of $\kappa_{xx}^\textrm{sp}$ in $\kappa_{xx}$ at low temperatures.

The thermal conduction of spin is also given by $\kappa_{xx}^\textrm{sp}=C_\textrm{sp}v_\textrm{sp} \ell_\textrm{sp}/3$, where $C_\textrm{sp}$, $v_\textrm{sp}$, and $\ell_\textrm{sp}$ is the specific heat, the velocity and the mean free path of the spin excitations, respectively.
As shown in Fig.\,\ref{fig:Specifit_Heat} (a), the specific heat does not show a large field suppression at 10\,T compared to that observed in $\kappa_{xx}$, excluding a possibility of suppressing the number of the spin excitations by a field-induced gap.
Also, $\ell_\textrm{sp}$ is known to become longer under a magnetic field because spin fluctuations are suppressed under a magnetic field. Therefore, the field suppression of $\kappa_{xx}^\textrm{sp}$ is caused by a field suppression effect on $v_\textrm{sp}$.
A similar field suppression effect on $v_\textrm{sp}$ has been shown in the 1D spin-chain compound~\cite{Sologubenko2007}.
Compared to the 1D spin-chain case, where the elementary excitations are well understood by the Bethe ansatz, the spin excitations in a spin liquid state of a 2D kagome is an extremely non-trivial issue. However, from the very similar field suppression effect on $v_\textrm{sp}$, we suggest a presence of a similar field suppression effect in Cd-K as that in the 1D spin-chain case.

Here, we consider the hump-like increase of $\kappa_{xx}$ observed near $T_{\text{N}}$. This increase is caused by the increase of $\kappa_{xx}^\textrm{ph}$ by a reduction of spin fluctuations~\cite{Hentrich2018} and/or the appearance of a magnon contribution in the ordered state\,\cite{YamashitaJPCM2020}. 
In the former case, as observed in $\alpha$-RuCl$_3$ (Ref.~\cite{Hentrich2018}), the increase of  $\kappa_{xx}$ at $T_{\text{N}}$ should be larger under higher fields because the spin fluctuations are more strongly suppressed under higher fields.
However, as shown in Fig.\,\ref{fig:kxx_vs_T}, the increase becomes smaller at higher fields.
This is consistent with the field suppression effect on $\kappa_{xx}^\textrm{sp}$. 
In addition, the large field suppression of $\kappa_{xx}$ at low temperatures suggests a dominant contribution of $\kappa_{xx}^\textrm{sp}$.
Therefore, the increase of $\kappa_{xx}$ below $T_{\text{N}}$ is likely attributed to the magnon contribution. 
The increase of $\kappa_{xx}$ below $T_{\text{N}}$ was observed larger in a better crystal with a larger $\kappa_{xx}$. We note that a similar sample dependence of magnon thermal conduction has been observed in volborthite\,\cite{YamashitaJPCM2020}, which also supports the presence of a magnon contribution below $T_{\text{N}}$.

A new field-induced peak is observed in the magnetic field dependence of $\kappa_{xx}$ around 7\,T below 0.3 K (Fig.\,\ref{fig:knrm_vs_B} (c)). 
The resonance scattering effect on phonons is excluded to explain the magnetic field dependence because the resonance scattering effect at 0.3 K is saturated above $\sim 2$\,T.
Also, an increase of $\kappa_{xx}^\textrm{ph}$ by suppressing the AFM phase can be excluded because $T_\textrm{N}$ does not depend on the field up to 10\,T (Fig.\,\ref{fig:Specifit_Heat}(a)).
The magnon contribution, which is observed as the hump-like increase in $\kappa_{xx}$ below $T_\textrm{N}$, is also excluded for the field-induced increase because the magnon contribution is suppressed by fields (Fig.\,\ref{fig:kxx_vs_T}).
Therefore, this increase of $\kappa_{xx}$ around 7\,T indicates an appearance of some field-induced spin excitations by closing a spin gap.
As shown in Fig.\,\ref{fig:Specifit_Heat}(c), the increase of $C_\textrm{sp}$ at 0.5\,K is also observed around 7\,T. This also supports the appearance of the field-induced spin excitations. Further, in the ordered phase, a finite thermal Hall effect is observed only above 7\,T (Fig.\,\ref{fig:dTy_vs_B} (c)), implying that the thermal Hall effect is caused by the field-induced spin excitations observed in the field dependence of $\kappa_{xx}$ and $C$.

	One possible origin of this spin gap is an anisotropy of the interactions. In fact, the energy scale caused by $J_d / J = 0.18$ is comparable to 7\,T.
	The temperature dependence of the threshold field of $\Delta T_{y}^\textrm{Asym}$ (grey circles in Fig.\,\ref{fig:PhaseDiag}) further supports that the emergence of the field-induced excitations is related to the closing of the spin gap.

\subsection{Thermal Hall Conductivity} \label{Dis:kxy} 

\subsubsection{Failure of the spin-only model for Cd-K} \label{Dis:kxy:SBMFT}

We first discuss the temperature dependence of $\kappa_{xy}/TB$ of all three kagome compounds of Cd-K, Ca-K~\cite{DokiPRL2018}, and volborthite~\cite{WatanabePNAS2016} in terms of the spin thermal Hall effect calculated by the SBMFT\,\cite{LeePRB2015, HanJPSJ2017}.
As shown in Fig.\,\ref{fig:kxy_vs_T}(b), $\kappa_{xy}/TB$ of these kagome antiferromagnets shows a similar temperature dependence.
As reported in Ref.\,\onlinecite{DokiPRL2018}, both the temperature dependence and the magnitude of $\kappa_{xy}/TB$ of Ca-K and volborthite show a good agreement with a simulation based on the SBMFT.
In the SBMFT framework, the kagome Heisenberg Hamiltonian with a Zeeman term and a DM interaction is diagonalized by taking a mean-field value of the bond operator of Schwinger bosons. From the energy bands and the Berry curvature calculated by the SBMFT, $\kappa_{xy}^\textrm{SBMFT}$ is calculated by Eq.\,\ref{eq_kxy} and is expressed by a dimensionless function $f_\textrm{SBMFT}$ as
\begin{eqnarray}
\label{eq:SBMFT}
\frac{\kappa_{xy}^\textrm{SBMFT}}{T}=\frac{k_\text{B}^2}{\hbar}\frac{Dg\mu_{\text{B}}B}{J^2}  f_\textrm{SBMFT}\left(\frac{k_{\text{B}}T}{J}\right).
\end{eqnarray}
To compare this SBMFT calculation, the thermal Hall conductivity per one 2D kagome layer is estimated from the experimental data by $\kappa_{xy}^\textrm{2D}=\kappa_{xy}d$, where $d$ ($d=7.0328$\,{\AA} for Cd-K\,\cite{OkumaPRB2017}) is the distance between the kagome layers.
We then compare $\kappa_{xy}^\textrm{2D}$ with $f_\textrm{SBMFT}$ by normalizing $\kappa_{xy}^\textrm{2D}$ as
\begin{eqnarray}
\frac{\kappa_{xy}^\textrm{2D}}{T}=\frac{k_\text{B}^2}{\hbar}\frac{Dg\mu_{\text{B}}B}{J^2}  f_\textrm{exp},
\end{eqnarray}
where $J$ and $D$ are the fitting parameters.

\begin{figure}[hbt]
	\includegraphics[width=8.6cm]{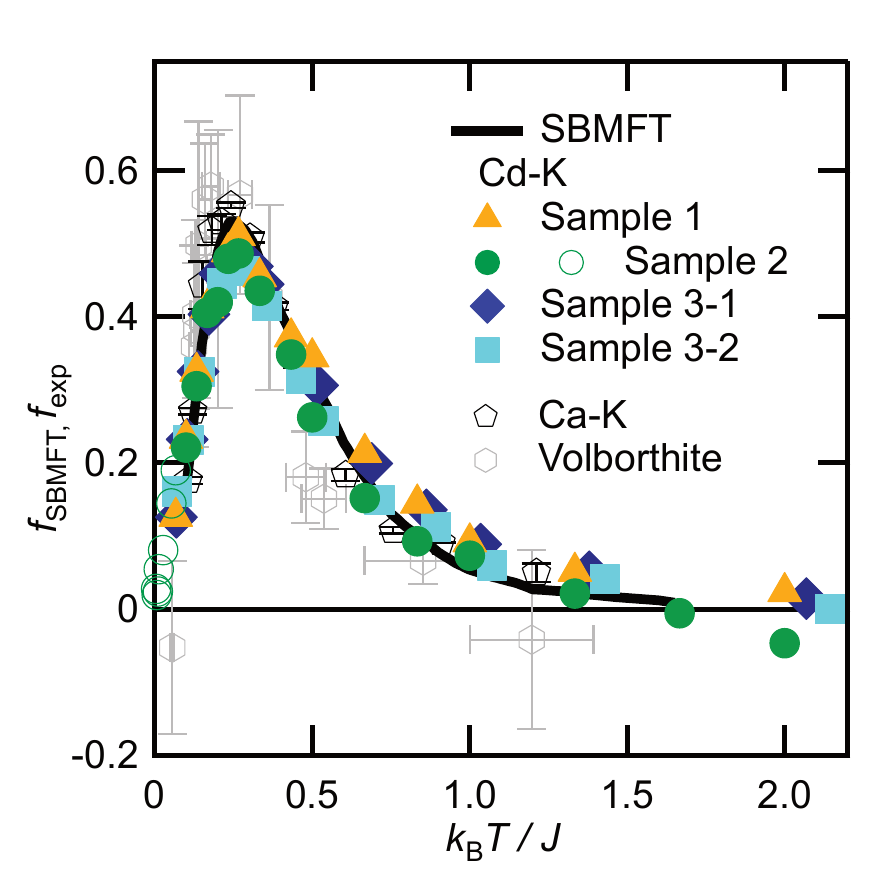}
	\caption{
		Normalized thermal Hall conductivity $f_\textrm{exp}$ of kagome lattice antiferromagnets fitted by the parameters listed in Table\,\ref{table1}. The solid line shows a numerical calculation of $f_\textrm{SBMFT}$ at $D/J=0.1$ by the Schwinger-boson mean field theory (SBMFT)\,\cite{DokiPRL2018}.
		The data of Ca-K and that of volborthite is taken from Ref.\,\onlinecite{DokiPRL2018} and Ref.\,\onlinecite{WatanabePNAS2016}, respectively.
	}
	\label{fig:SBMFT_fitting}
\end{figure}

\begin{table}[hbt]
	\caption{
		Values of $J$ and $\vert D/J\vert$ used to fit $\kappa_{xy}^\textrm{2D}$ to the SBMFT simulation (Fig.\,\ref{fig:SBMFT_fitting}) for kagome lattice antiferromagnets. The data of Ca-K and that of volborthite is taken from Ref.\,\onlinecite{DokiPRL2018} and Ref.\,\onlinecite{WatanabePNAS2016}, respectively.
	}
	\label{table1}
	\begin{ruledtabular}
		\begin{tabular}{cccd}
			\multicolumn{1}{c||}{Material}&
			\multicolumn{1}{c}{Sample No.}&
			\multicolumn{1}{c}{$J/k_{\text{B}}$} (K)&
			\multicolumn{1}{c}{$D/J$}\\
			%\mbox{Three}&\mbox{Four}&\mbox{Five}\\
			\hline
			\multicolumn{1}{c||}{Cd-Kapellasite}&1&30&0.28\\
			\multicolumn{1}{c||}{ }&2&30&0.09\\
			\multicolumn{1}{c||}{ }&3-1&29&0.65\\
			\multicolumn{1}{c||}{ }&3-2&28&0.6\\
			\multicolumn{1}{c||}{Ca-Kapellasite\,\cite{DokiPRL2018}}& &66&0.12\\
			\multicolumn{1}{c||}{Volborthite\,\cite{WatanabePNAS2016}}& &60&-0.07\\
		\end{tabular}
	\end{ruledtabular}
\end{table}

	By adopting this SBMFT analysis, we fitted $\kappa_{xy}^\textrm{2D}$ of Cd-K by tuning the fitting parameters of $J$ and $D$.
	Although all the $\kappa_{xy}^\textrm{2D}$ data of Cd-K well converges to one single curve given by the SBMFT (solid line in Fig.\,\ref{fig:SBMFT_fitting}) by  the fitting parameters listed in Table \ref{table1}, these fittings result in \emph{unphysical} fitting parameters for Cd-K.
	First, $J=30$\,K used for the fit of Cd-K is considerably smaller than that estimated by the temperature dependence of $\chi$ ($J=45$\,K)~\cite{OkumaNComms2019}.
More importantly, the magnitude of $D$ used to the fit of $\kappa_{xy}$ of Cd-K differs in a factor of 7 among the Cd-K samples owing to the very different magnitudes of $\kappa_{xy}$.
This large difference of $D$ in Cd-K samples is too large to explain it by the ambiguity in estimating the sample dimensions.
In addition, the largest value of $D/J=0.65$ is more than three times larger than the value of $D/J\sim0.19$  estimated from the deviation of the $g$ factor from 2~\cite{OkumaNComms2019}, 
This is in sharp contrast to the analysis done for Ca-K\,\cite{DokiPRL2018} in which both $J$ and $D$ determined by the SBMFT fit of $\kappa_{xy}$ well coincide with the value estimated from the temperature dependence of $\chi$ and that from the deviation of the $g$ factor, respectively.
These results indicate that the origin of the thermal Hall effect in Cd-K is different from the spin thermal Hall effect observed in Ca-K\,\cite{DokiPRL2018}.

\subsubsection{Phonon thermal Hall effect in Cd-K} \label{Dis:kxy:phonon}

As discussed in section\,\ref{Dis:kxx}, $\kappa_{xx}$ of Cd-K is given by a sum of $\kappa_{xx}^\textrm{ph}$ and $\kappa_{xx}^\textrm{sp}$. 
Therefore, $\kappa_{xy}$ of Cd-K can also contain a phonon contribution $\kappa_{xy}^\textrm{ph}$ in addition to a spin contribution $\kappa_{xy}^\textrm{sp}$.

Thermal Hall effects of phonons has been reported in various compounds~\cite{Strohm2005,Sugii2017,Hirokane2019,XiaokangLi2020,Grissonnanche2020}.
In the nonmagnetic insulator SrTiO$_3$, in which only phonons are responsible for the thermal transport, 
$\kappa_{xy}$ of phonons is found to show a peak at the same temperature of the peak in $\kappa_{xx}$.
We thus checked this relation for Cd-K.
In Cd-K, $\kappa_{xx}$ of Cd-K contains a spin contribution $\kappa_{xx}^\textrm{sp}$ which becomes dominant at lower temperatures.
On the other hand, as shown in Fig.\,\ref{fig:knrm_vs_B}, a magnetic field suppresses a large portion of $\kappa_{xx}^\textrm{sp}$ whereas it slightly increases $\kappa_{xx}^\textrm{ph}$.
Therefore, $\kappa_{xx}^\textrm{ph}$ can be estimated by $\kappa_{xx}$ at 15\,T.

\begin{figure*}[h!bt]
	\includegraphics[width=\linewidth]{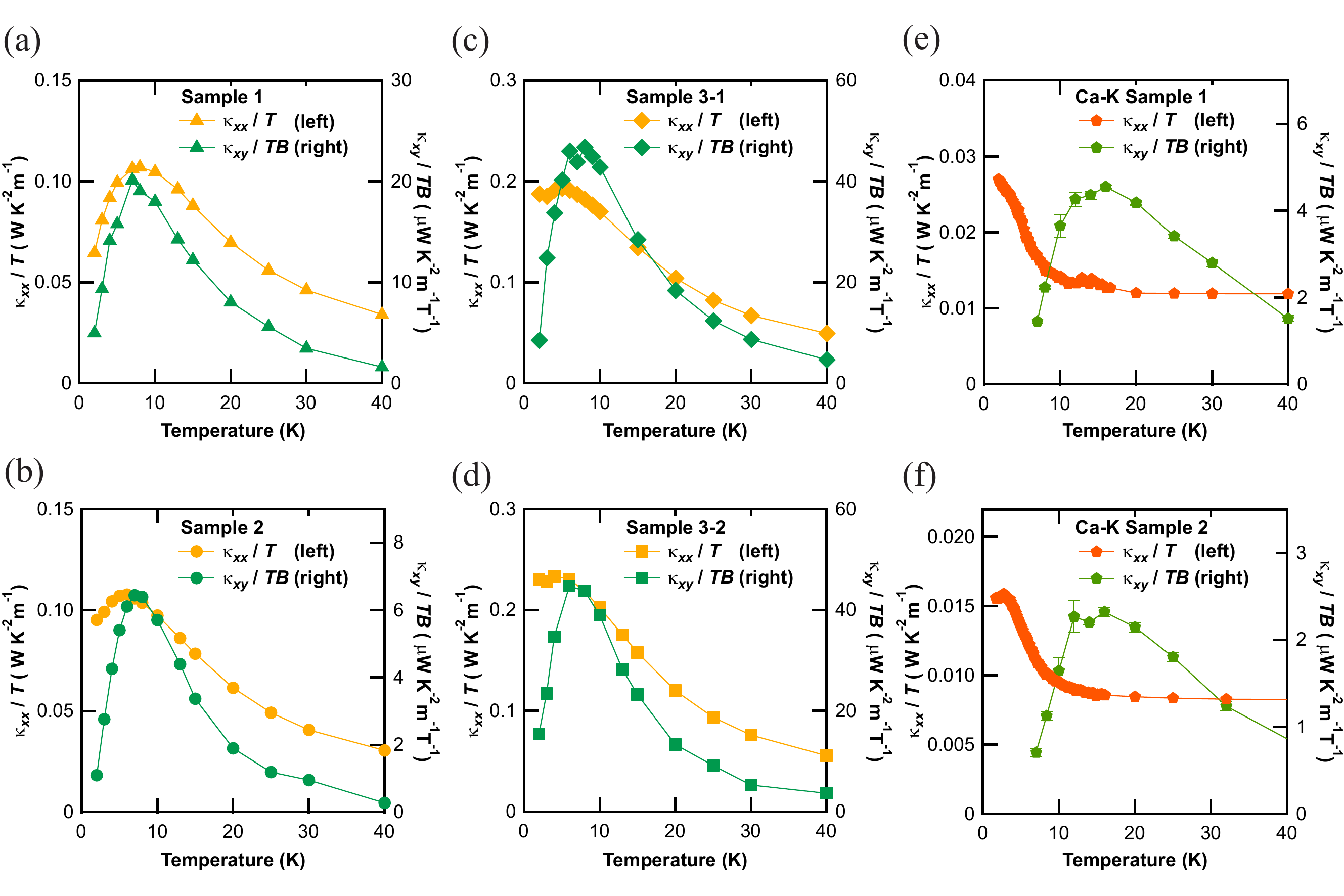}
	\caption{\label{fig:kxy_vs_kxx15T}
		Temperature dependence of $\kappa_{xx}/T$ (left axis) and that of $\kappa_{xy}/TB$ (right axis) at 15\,T of all Cd-K samples (a--d), and Ca-K (e, f). The data of Ca-K is taken from Ref.\,\onlinecite{DokiPRL2018}.
	}
\end{figure*}

Figure\,\ref{fig:kxy_vs_kxx15T} shows the temperature dependence of $\kappa_{xx}/T$ (left axis) and that of $\kappa_{xy}/TB$ (right axis)  at 15\,T of all Cd-K  samples, together with those of Ca-K~\cite{DokiPRL2018}.
As shown in Figs.\,\ref{fig:kxy_vs_kxx15T}(a--d), $\kappa_{xx}/T$ at 15\,T shows a peak at almost the same temperature of the peak of $\kappa_{xy}/TB$, which resembles the case of the phonon thermal Hall effect observed in SrTiO$_3$ (Ref.~\cite{XiaokangLi2020}).
\textcolor{black}{
	Further, the maximum of $\kappa_{xy}/TB$ almost linearly increases with $\kappa_{xx}/T$, resulting in the estimation of the Hall angle ($\kappa_{xy}/\kappa_{xx}$) at the peak temperature as $\sim 4\times 10^{-3}$ at 15\,T. This Hall angle is close to that observed in the phonon thermal Hall effects in SrTiO$_3$~\cite{XiaokangLi2020} and cuprates~\cite{Grissonnanche2020}, supporting the phonon origin of $\kappa_{xy}$ in Cd-K at 15\,T.
	Therefore, we conclude that $\kappa_{xy}$ of Cd-K at 15\,T contains a dominant phonon contribution.
}

We also find that this is clearly not the case for Ca-K~\cite{DokiPRL2018}. As shown in Figs.\,\ref{fig:kxy_vs_kxx15T}(e) and (f), $\kappa_{xx}/T$ at 15\,T peaks at a much lower temperature than that of $\kappa_{xy}/TB$, which is consistent with the spin origin of $\kappa_{xy}$ in Ca-K.

\textcolor{black}{
	\subsubsection{Spin thermal Hall effect in Cd-K} \label{Dis:kxy:spin}
}

The energy scale of the phonon thermal Hall effect should be given by the Debye temperature, which is estimated as 220\,K for Cd-K from the temperature dependence of the specific heat at high temperatures~\cite{SM}.
Since this energy scale is order of magnitude larger than that of magnetic field of 15\,T,  $\kappa_{xy}^\textrm{ph}$ is expected to have a linear field dependence, which is indeed
 confirmed in the field dependence of $\kappa_{xy}$ at high temperatures (Fig.\,\ref{fig:k_xy_vs_B_all_samples}).
On the other hand,  the field dependence of $\kappa_{xy}$ becomes non-linear as lowering temperature. As shown in Fig.\,\ref{fig:k_xy_vs_B_all_samples}, the slope of $\kappa_{xy}/B$ below 20\,K becomes larger at lower fields, suggesting an emergence of an additional thermal Hall effect.
In this temperature range, $\kappa_{xx}$ also starts to show the field suppression effect as discussed in section\,\ref{Dis:kxx}, which is given by the field suppression effect on $\kappa_{xx}^\textrm{sp}$.
Therefore, this non-linear field dependence of $\kappa_{xy}$  at low temperatures suggests an emergence of a spin contribution $\kappa_{xy}^\textrm{sp}$ which is related to $\kappa_{xx}^\textrm{sp}$.

We estimate this additional spin component $\delta \kappa_{xy}(B)$ at each temperature by
\begin{eqnarray}
	\delta \kappa_{xy}(B)=\kappa_{xy}(B) - \frac{B}{15} \kappa_{xy}(15 \,\textrm{T}).
\end{eqnarray}
Note that $\delta \kappa_{xy}(B)$ gives a lower bound of  $\kappa_{xy}^\textrm{sp}$ because the spin contribution in $\kappa_{xy}$ is not fully quenched at 15\,T.

\begin{figure*}[h!bt]
	\includegraphics[width=8.6cm]{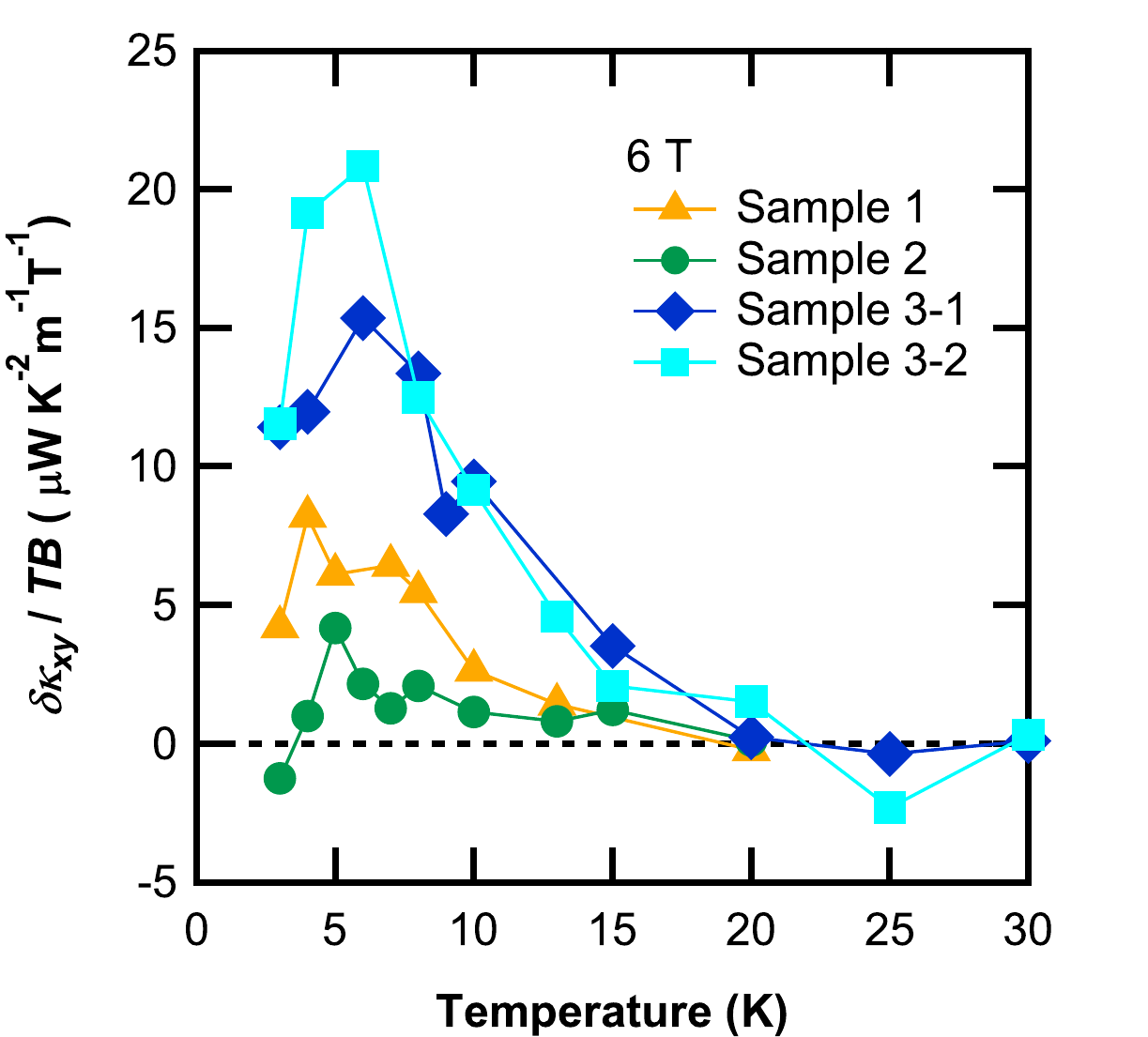}
	\caption{\label{fig:dkxy_T}
		Temperature dependence of $\delta \kappa_{xy}/TB$ at 6\,T of all Cd-K samples.
	}
\end{figure*}

We investigate the temperature dependence  of $\delta \kappa_{xy}/TB$ at 6\,T (Fig.\,\ref{fig:dkxy_T}) where the field dependence of  $\delta \kappa_{xy}(B)$ shows a peak (see Fig.\,\ref{fig:k_xy_vs_B_all_samples}). 
As shown in Fig.\,\ref{fig:dkxy_T}, $\delta \kappa_{xy}/TB$ of all the Cd-K samples starts to appear below $\sim$25\,K, much lower than that of $\kappa_{xy}^\textrm{ph}/TB$  at 15\,T which persists up to $\sim$60\,K (Fig.\,\ref{fig:kxy_vs_T}).
This appearance of  $\delta \kappa_{xy}/TB$ coincides with the field suppression effect on $\kappa_{xx}^\textrm{sp}$ (Fig.\,\ref{fig:kxx_vs_T}), indicating that the spin contribution $\kappa_{xy}^\textrm{sp}$ is given by $\delta \kappa_{xy}/TB$.

\begin{figure*}[h!bt]
	\includegraphics[width=8.6cm]{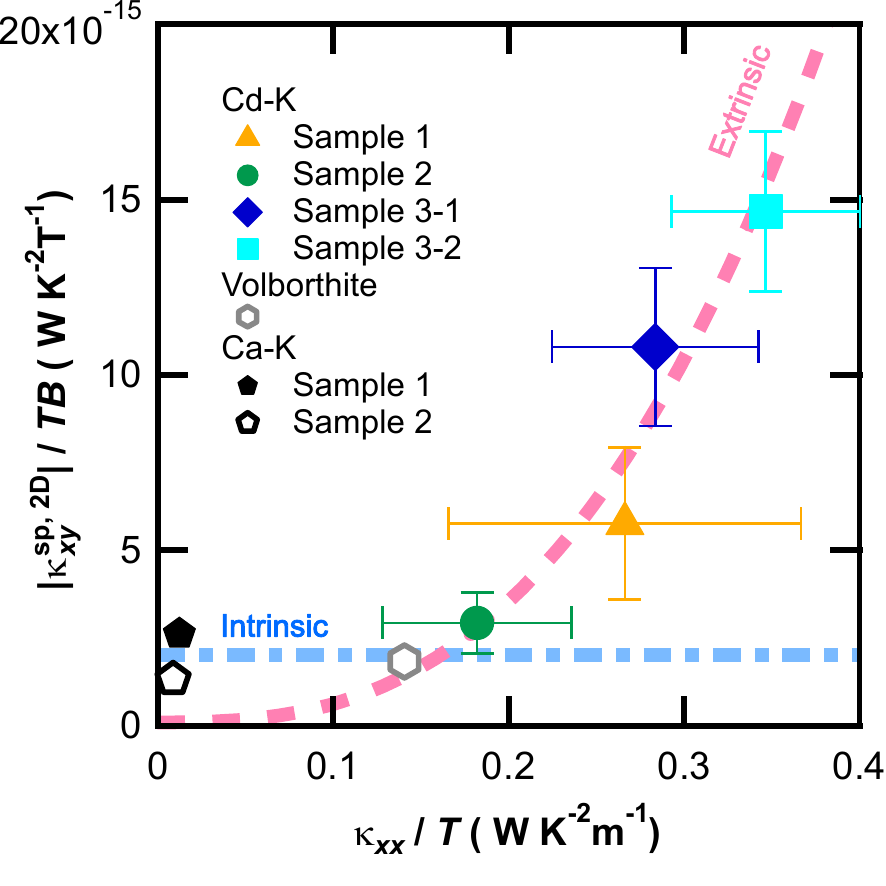}
	\caption{\label{fig:kxy_sp_2D_vs_kxx}
		The spin thermal Hall conductivity per the 2D kagome layer
		$\left| \kappa_{xy}^\textrm{sp, 2D}  \right| /TB$ of Cd-K, Ca-K \cite{DokiPRL2018}, and volborthite~\cite{WatanabePNAS2016} 
		plotted as a function of the longitudinal thermal conductivity $\kappa_{xx}/T$ at 0 T. The value of $\kappa_{xx}/T$ is taken at the peak temperature of $\delta \kappa_{xy}/TB$ for Cd-K (4--6\,K, see Fig.\,\ref{fig:dkxy_T}) and that of $\left| \kappa_{xy}^\textrm{2D}  \right| /TB$ for Ca-K and volborthite (16\,K).
		The error bars represent the ambiguity in estimating the absolute vlue of $\kappa_{xx}$ and $\kappa_{xy}$ owing to the non-rectangular shape of the samples (see SM~\cite{SM} for details).
		The blue and pink dashed lines are guides to the eye for the intrinsic and extrinsic contributions, respectively (see the main text).
	}
\end{figure*}

Figure\,\ref{fig:dkxy_T} shows that the magnitude of $\delta \kappa_{xy}/TB$ depends on the sample, implying that the magnitude of $\kappa_{xy}^\textrm{sp}$ depends on $\kappa_{xx}^\textrm{sp}$.
As shown by  the field suppression effect on $\kappa_{xx}^\textrm{sp}$ (Fig.\,\ref{fig:kxx_vs_T}), $\kappa_{xx}$ at the peak  temperature of $\delta \kappa_{xy}/TB$ contains a significant spin contribution at 0\,T.
We therefore investigate the dependence of the maximum of $\delta \kappa_{xy}/TB$ on $\kappa_{xx}/T$ at 0 T at the peak temperature of $\delta \kappa_{xy}/TB$ for each Cd-K samples (4--6\,K).
We also check this relation for Ca-K~\cite{DokiPRL2018} and volborthite~\cite{WatanabePNAS2016}.
To compare the results in different samples, we estimate $\kappa_{xy}^\textrm{sp, 2D} = \kappa_{xy}^\textrm{sp} \times d$ and assemble all data in Fig.\,\ref{fig:kxy_sp_2D_vs_kxx} by plotting $\left| \kappa_{xy}^\textrm{sp, 2D}  \right| /TB$ as a function of $\kappa_{xx}/T$.

Most remarkably, as shown in Fig.\,\ref{fig:kxy_sp_2D_vs_kxx}, $\left| \kappa_{xy}^\textrm{sp, 2D}  \right|$ data of three different kagome materials appears to show a smooth function of $\kappa_{xx}/T$;  $\left| \kappa_{xy}^\textrm{sp, 2D}  \right|$ stays a constant in the low-$\kappa_{xx}$ region whereas $\left| \kappa_{xy}^\textrm{sp, 2D}  \right|$ increases as $\kappa_{xx}$ does.
Given that all these compounds share a similar kagome structure (especially for Ca-K and Cd-K), $\kappa_{xx}$ ($\propto C v^2 \tau$) directly reflects the scattering time $\tau$ determined by disorder scattering.
Therefore, Fig.\,\ref{fig:kxy_sp_2D_vs_kxx} shows $\tau$ dependence of $\left| \kappa_{xy}^\textrm{sp, 2D}  \right|$.

As discussed in Ref.\,\cite{DokiPRL2018}, $\left| \kappa_{xy}^\textrm{sp, 2D}  \right| /TB$ of Ca-K and volborthite can be well explained by the SBMFT calculation $\kappa_{xy}^\textrm{SBMFT}/TB$ of which the magnitude is only given by $D/J$ and does not depend on $\kappa_{xx}$.
We denote this as the \emph{``intrinsic"} contribution (the blue dashed line in Fig.\,\ref{fig:kxy_sp_2D_vs_kxx}).
On the other hand, $\left| \kappa_{xy}^\textrm{sp, 2D}  \right| /TB$ of Cd-K samples clearly exceeds the intrinsic contribution and increases as $\kappa_{xx}$ does, which should be denoted as an \emph{``extrinsic"} contribution (the pink dashed line in Fig.\,\ref{fig:kxy_sp_2D_vs_kxx}).

This $\kappa_{xx}$ dependence of  $\left| \kappa_{xy}^\textrm{sp, 2D}  \right| /TB$ bears similarity to that of the anomalous Hall effect (AHE) in ferromagnetic metals.
In the AHE, it has been known that the dominant mechanism of AHE depends on the magnitude of the longitudinal conductivity\,\cite{OnodaPRL2006, OnodaPRB2008}; the intrinsic mechanism by the Berry curvature of the energy bands is dominant in a moderate dirty metal whereas the extrinsic mechanism by skew scatterings is dominant for a super-clean metal. 
This good analogy between the spin thermal Hall effect in the kagome materials and the AHE in ferromagnetic metals indicates a presence of a similar duality of intrinsic-extrinsic mechanisms for the spin thermal Hall effect of an insulator.

\subsubsection{Spin-phonon coupling and re-emergence of $\kappa_{xy}^\textrm{ph}$ in the AFM phase} \label{Dis:kxy:spinphonon}

In the AFM phase of Cd-K, no thermal Hall effect is observed below $\sim$7\,T (Fig.\,\ref{fig:dTy_vs_B}(c)), showing that both $\kappa_{xy}^\textrm{ph}$ and $\kappa_{xy}^\textrm{sp}$ disappear in the AFM phase at low fields.
This absence of  $\kappa_{xy}^\textrm{sp}$ in the AFM phase is consistent with the theoretical prediction~\cite{Mook2019}.
On the other hand, $\kappa_{xy}/TB$ in the AFM phase re-emerges  above $\sim$7\,T (Fig.\,\ref{fig:dTy_vs_B}(c)).
The temperature dependence of $\kappa_{xy}/TB$ at 14\,T below $T_\textrm{N}$ well follows that at the spin liquid phase at 15\,T in which the phonon contribution is dominant (see the inset of Fig.\,\ref{fig:kxy_vs_T}(b)).
\textcolor{black}{
	This smooth connection of $\kappa_{xy}/TB$ above and below $T_\textrm{N}$ at high fields implies that this high-field $\kappa_{xy}$ below $T_\textrm{N}$ is mostly given by phonons.
	This is also consistent with that $\kappa_{xy}^\textrm{sp}$ above $T_\textrm{N}$ is observed at low fields, and the strong field suppression effect on $\kappa_{xx}^\textrm{sp}$.
}
Therefore, the appearance of $\kappa_{xy}$ in the AFM phase above $\sim 7$\,T (Fig.\,\ref{fig:dTy_vs_B}(c) and Fig.\,\ref{fig:dTy_vs_B_allS}) shows a re-emergence of $\kappa_{xy}^\textrm{ph}$ which disappears at low fields.
This absence of $\kappa_{xy}^\textrm{ph}$ in the low-field AFM phase demonstrates that the phonons cannot alone exhibit the thermal Hall effect, putting a strong constraint on the origin of the phonon Hall effect.
In other words, $\kappa_{xy}^\textrm{ph}$ needs to be triggered by the field-induced excitations observed in the field dependence of $\kappa_{xx}$ (Fig.\,\ref{fig:knrm_vs_B}(c)) and that of $C$ (Fig.\,\ref{fig:Specifit_Heat}(b)) through a spin-phonon coupling.

Thermal Hall effects by phonons have been theoretically studied with respect to various aspects~\cite{Sheng2006,Kagan2008,WangZhang2009,LifaZhang2010,Qin2012,MoriPRL2014,Saito2019,Zhang2019,ChenKivelsonSun2020}.
The absence of the stand-alone phonon thermal Hall effect in Cd-K is inconsistent with an intrinsic mechanism, but rather points to extrinsic origins with microscopic couplings between phonons and the field-induced excitations.
Such microscopic coupling has also been suggested to play an important role in the thermal Hall effect observed in multiferroics~\cite{Ideue2017} where a large thermal Hall effect is observed in the ferrimagnetic phase despite the absence of the conventional magnon Hall effect.
Also, it has been pointed out that a magnon-phonon coupling induces a thermal Hall effect even in the system where neither phonons nor magnons alone show a thermal Hall effect~\cite{Zhang2019}.
Therefore, the absence and the presence of $\kappa_{xy}^\textrm{ph}$ in the AFM phase of Cd-K suggests that the phonon thermal Hall effect in Cd-K has an extrinsic origin requiring a spin-phonon coupling with the field-induced spin excitations.
At present, the details of the magnetic structure of Cd-K have not been known.
Further studies, including NMR or neutron scattering experiments to clarify the magnetic excitations in the AFM phase under the low and the high magnetic fields, will be important to reveal the origins of the thermal Hall effects in Cd-K.

One clearly has to wonder why, despite the dual origin (spins and phonons) of the thermal Hall effects in the Cd-K compounds, the scaling fit derived entirely from the SBMFT works so well as shown in Fig.\,\ref{fig:SBMFT_fitting}. 
This is not an unreasonable conclusion, however, provided we further assume that $\kappa_{xy}^\textrm{sp}$ is proportional to $\kappa_{xy}^\textrm{ph}$, due to the fact that the two excitations are microscopically coupled.

	We note that the field dependence of the thermal Hall effect in the ordered phase of Cd-K is in sharp contrast to that observed in Ca-K.
	In the ordered phase in Ca-K, a finite $\kappa_{xy}$ is observed only in a low-field and is absent above $\sim 6$ T\,\cite{DokiPRL2018}, whereas the ordered state in Ca-K is suggested to be the same $q = 0$ negative chiral state\,\cite{Ihara2020, Iida2020}.
	In the two compounds, the magnitude of $\kappa_{xx}$ is quite different (Fig.\,\ref{fig:kxx_vs_T})) owing to the different crystal quality. Moreover, according to the theoretical study~\cite{Mook2019}, $\kappa_{xy}^\textrm{sp}$ in the negative chiral state has been suggested to depend on the rotation angle of the spin in the kagome plane.
	Therefore, this different field dependence of $\kappa_{xy}$ could be attributed to the different magnitude of $\kappa_{xx}$ and the spin angle in the AFM state (see SM~\cite{SM} for details).

\section{\label{sec:summary}Summary}

We have investigated $\kappa_{xx}$ and $\kappa_{xy}$ of three Cd-K samples. From the field suppression effect on $\kappa_{xx}$, we find that the spin contribution $\kappa_{xx}^\textrm{sp}$ sets in $\kappa_{xx}$ below $\sim$25\,K, which becomes dominant at lower temperatures.
Below $T_{\text{N}}$, we find a new peak in the field dependence of $\kappa_{xx}$ at $\sim7$ T. Above 7\,T, we also find a field-induced increase both in the specific heat and in the thermal Hall effect.

\textcolor{black}{
	Clear thermal Hall effects have been observed in the spin liquid states of all Cd-K samples. 
	We find that  $\kappa_{xy}$ of all Cd-K samples shows a virtually identical temperature dependence with a peak at almost the same temperature.
	On the other hand, the magnitude of $\kappa_{xy}$ of samples with high $\kappa_{xx}$ substantially exceeds that expected for the intrinsic spin thermal Hall effect determined by $D/J$ in the SBMFT framework.
	We conclude that a phonon thermal Hall  $\kappa_{xy}^\textrm{ph}$ contributes to $\kappa_{xy}$ of all Cd-K, deduced from the similar temperature dependence of $\kappa_{xy}$ and $\kappa_{xx}$ at 15\,T and the positive correlation between them (Fig.\,\ref{fig:kxy_vs_kxx15T}).
}

\textcolor{black}{
	We further find that the non-linear field dependence of $\kappa_{xy}$ at low temperatures (Fig.\,\ref{fig:k_xy_vs_B_all_samples}) shows the emergence of a spin thermal Hall effect $\kappa_{xy}^\textrm{sp}$ at low fields (Fig.\,\ref{fig:dkxy_T}).
	Moreover, the appearance of $\kappa_{xy}^\textrm{sp}$ coincides in its temperature range with the region where the field suppression effect on $\kappa_{xx}^\textrm{sp}$ is being observed (Fig.\,\ref{fig:kxx_vs_T}).
	Most remarkably, whereas $\kappa_{xy}^\textrm{sp}$ does not depend on $\kappa_{xx}$ and is well explained by the intrinsic thermal Hall effect given by the SBMFT framework in the low-$\kappa_{xx}$ region,
	$\kappa_{xy}^\textrm{sp}$ exceeds the intrinsic contribution with a positive correlation to  $\kappa_{xx}$ for the high-$\kappa_{xx}$ region (Fig.\,\ref{fig:kxy_sp_2D_vs_kxx}).
	This $\kappa_{xx}$ dependence of $\kappa_{xy}^\textrm{sp}$ points to the presence of both intrinsic and extrinsic mechanisms for the spin thermal Hall effect in the kagome materials, as was the case for the anomalous Hall effects in ferromagnetic metals.
}

In addition, we find that both  $\kappa_{xy}^\textrm{ph}$ and  $\kappa_{xy}^\textrm{sp}$ disappear in the AFM phase at low fields. At high fields above 7\,T, we find that $\kappa_{xy}^\textrm{ph}$ is induced concomitantly with the field-induced spin excitations observed in the field dependence of $\kappa_{xx}$ and $C$.
These results suggest that field-induced spin excitations give rise to a recovery of the phonon thermal Hall effect of Cd-K. 
We conclude that the phonons alone do not exhibit the thermal Hall effect and require to merge with the field-induced spin excitations through a spin-phonon coupling to appear.

\begin{acknowledgments}
	This work was supported by Grants-in-Aid for Scientific Research (KAKENHI) (No.\,19H01809, No.\,19H01848, and No.\,19K21842).
\end{acknowledgments}

	% MY: 19H01848,19K21842 
	% NK: 19H01809 

%\bibliography{bibv1}

\begin{thebibliography}{64}
\bibitem [{FNd()}]{FNdecomposing}%
  \BibitemOpen
  \href@noop {} {}\bibinfo {note} {We note that decompising the thermal
  conductivity (for both $\kappa_{xx}$ and $\kappa_{xy}$) into two independent
  phonon and spin contribution is an assumption made for convenience for
  discussion. For a full understanding of the results, one needs to incorporate
  a merged contribution given by a spin-phonon coupling. However, given that
  the energy scale of a spin-phonon coupling in Cu-based compounds is usually
  smaller than that of phonons (the Debye temperature) or the spin interaction
  energy, this two-channel model is often a good approximation.}\BibitemShut
  {Stop}%
\end{thebibliography}

%%%%%%%%%%%%%%%%%%%%%%%%%%%%%%%%%%%%%%%%%%%%%%%
%%	arXiv	SM

\newpage

\renewcommand{\theequation}{S\arabic{equation}}
\setcounter{equation}{0}
\renewcommand{\thetable}{S\arabic{table}}
\renewcommand{\bibnumfmt}[1]{[S#1]}
\renewcommand{\citenumfont}[1]{S#1}
\setcounter{figure}{0}
\renewcommand{\thefigure}{S\arabic{figure}}

\begin{titlepage}
	\begin{center}
		\vspace*{12pt}
		{\Large Supplemental Material for ``Thermal Hall Effects of Spins and Phonons in Kagome Antiferromagnet Cd-Kapellasite"}
		\vspace{12pt} \\
		\begin{tabular}{c}
			Minoru Yamashita, Masaaki Shimozawa, Shunichiro Kittaka, Toshiro Sakakibara, \\
			Ryutaro Okuma, Zenji Hiroi, Hyun-Yong Lee, Naoki Kawashima, Jung Hoon Han, and Minoru Yamashita
		\end{tabular} \vspace{3pt} \\
	\end{center}
\end{titlepage}

\subsection{Errors in estimating the geometrical factors}

	Because of the non-rectangular shape of the sample,  the sample with $w$ is determined by the average width between the thermal contacts. This approximation has the largest ambiguity in estimating the geometrical factors, resulting in a typical error of 20--30\% depending on the sample shape.
	The ambiguity of about 5\% is also caused in estimating $L$ and $w'$ owing to the size of the thermal contacts.
	The thickness of the Cd-kapellasite samples is uniform, resulting in an error less than a few percent.
	In total, a typical ambiguity from the shape effect is up to about 40\% in estimating the absolute value of $\kappa_{xx}$ and $\kappa_{xy}$.

\subsection{Heater power dependence of the longitudinal and transverse temperature differences}\label{sec:Qdep}

\begin{figure}[h!b!]
	\includegraphics[width=0.5\linewidth]{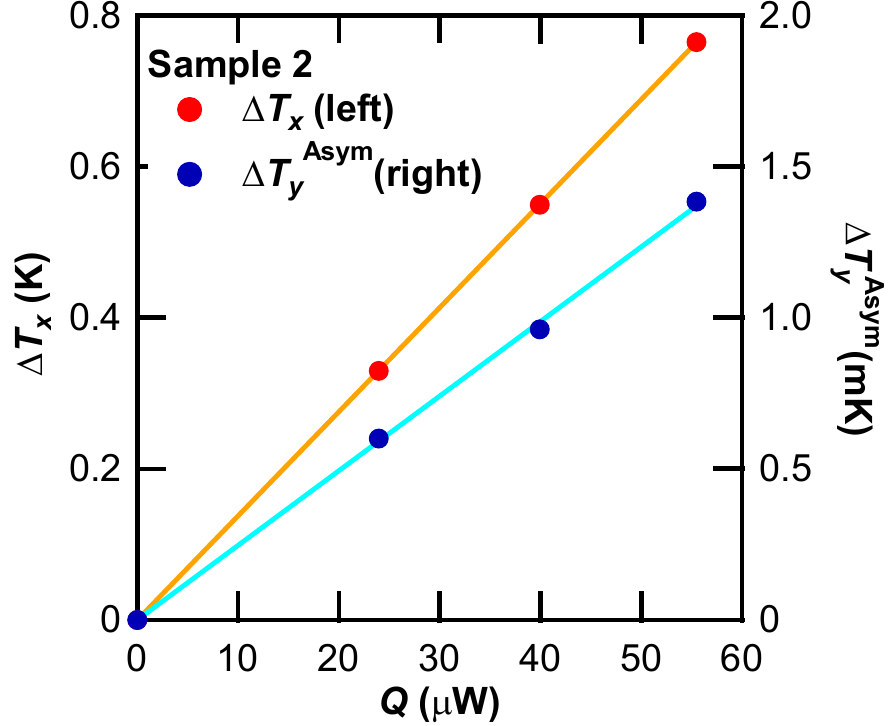}
	\caption{\label{fig:Qdep}
	The heat current $Q$ dependence of  $\Delta T_x$ and $\Delta T_y^\textrm{Asym}$ at 10\,K, $\pm 15$\,T. The standard error of the data was smaller than the symbol size.
	 }
\end{figure}

Figure \,\ref{fig:Qdep} shows the heat current ($Q$) dependence of the longitudinal ($\Delta T_x = T_\textrm{High} - T_\textrm{L1}$) and the antisymmetrized transverse ($\Delta T_y^\textrm{Asym}(B) = \left( \Delta T_y(+B) - \Delta T_y(-B) \right)/2$, where $\Delta T_y = T_\textrm{L1} - T_\textrm{L2}$) temperature difference at 10\,K, $\pm 15$\,T.
As shown in Fig.\,\ref{fig:Qdep}, both $\Delta T_x$ and $\Delta T_y^\textrm{Asym}$ shows a linear $Q$ dependence, validating the estimation of the longitudinal ($\kappa_{xx}$) and the transverse ($\kappa_{xy}$) thermal conductivity by Eq.\,(2) in the main text.

\subsection{Temperature stability of the thermal conductivity and the thermal Hall measurements}\label{sec:temp_stb}

During $\kappa_{xx}$ and $\kappa_{xy}$ measurements, the sample temperature estimated by $\left(T_\textrm{High} + T_\textrm{L1} \right)/2$
was stabilized by a PID control to the heater on the LiF heat bath.
This ensures that measurements at different $Q$ or at different magnetic fields were performed at the same sample temperature, resulting in a better S/N for the most sensitive $\kappa_{xy}$ measurements.

\begin{figure}[hb]
	\includegraphics[width=0.6\linewidth]{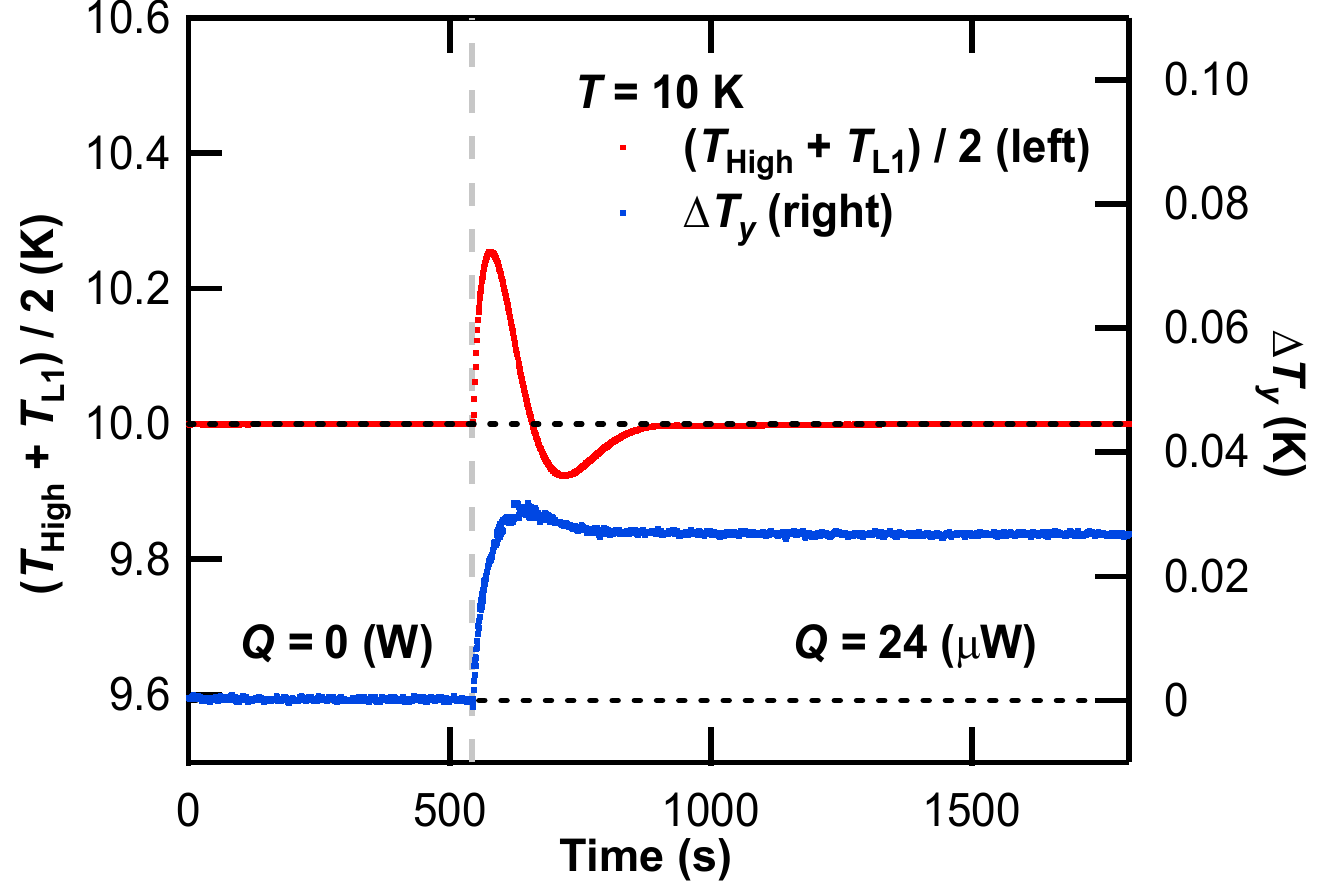}
	\caption{\label{fig:temp_stb}
		A representative time evolution of the sample temperature $\left(T_\textrm{High} + T_\textrm{L1} \right)/2$ and $\Delta T_y$ at 10\,K, +15\,T of Sample 2.
		A heat current $Q$ was turned on at the time shown by the dashed line.
	}
\end{figure}

Figure\,\ref{fig:temp_stb} shows a representative time evolution of the sample temperature and $\Delta T_y$ at 10\,K.
After stabilizing the sample temperature, we averaged the data for 360 sec for each measurement. The stability of the sample temperature during the period is $\pm 77$\,$\mu$K at 10\,K, which is small enough to resolve $\Delta T_y^\textrm{Asym}$ shown in Fig.\,\ref{fig:Qdep}.

\subsection{Estimation of the phonon specific heat}

\begin{figure}[!bth]
	\includegraphics[width=0.8\linewidth]{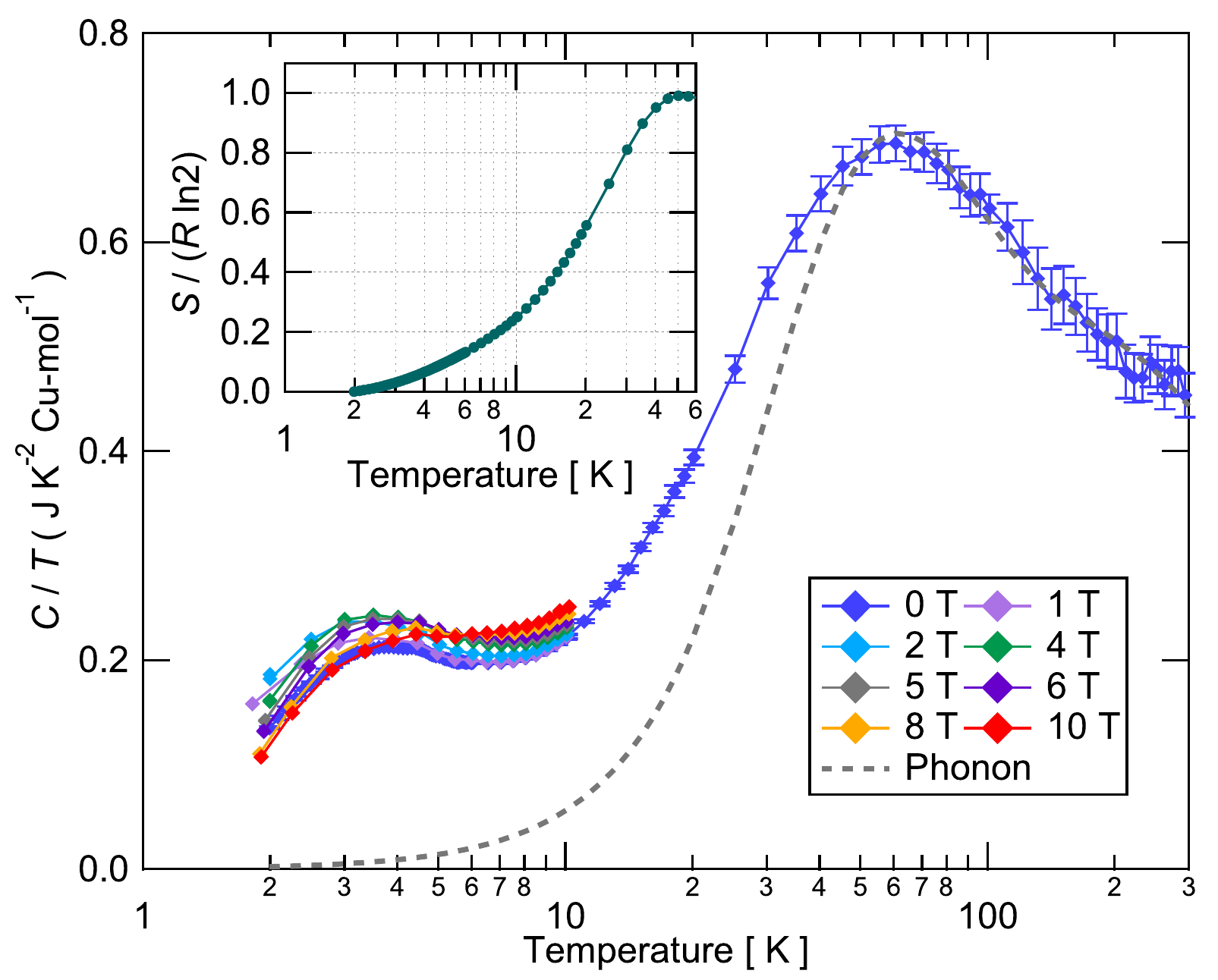}
	\caption{\label{fig:C_SM}
		The temperature dependence of the specific heat divided by the temperature ($C/T$) up to 300\,K.
		The dashed line shows the phonon contribution $C_\textrm{ph}$ estimated by the high temperature fitting of the data (($45 \le T \le 300$\,K) by Eq.\,\ref{eq:C_T_SM}.
		The error bars represent the standard deviation of the measurements, which are smaller than the symbol size below 10 K.
		The inset shows the temperature dependence of the magnetic entropy calculated by $C - C_\textrm{ph}$ at 0\,T.
	}
\end{figure}

Figure\,\ref{fig:C_SM} shows the temperature dependence of the specific heat divided by the temperature ($C/T$) up to 300\,K. 
Given the absence of a non-magnetic compound isostructural to Cd-K, we estimate the phonon contribution ($C_\textrm{ph}$) from a high-$T$ fitting of the data ($45 \le T \le 300$\,K) by a linear combination of Debye and Einstein functions as
\begin{eqnarray} \label{eq:C_T_SM}
C_\textrm{ph} &=& A_D \cdot 9R \left( \frac{T}{T_D} \right)^3 \int^{T_D/T}_0 \frac{x^4 e^x}{(e^x -1)^2}\textrm{d}x \nonumber \\
&&+A_E \cdot 3 R \left( \frac{T_E}{T} \right)^2 \frac{e^{T_E / T}}{\left( e^{T_E / T} -1 \right)^2}
\end{eqnarray}

As shown in the dashed line in Fig.\,\ref{fig:C_SM}, a good fitting is obtained by adjusting the fitting parameters ($A_D$, $T_D$, $A_E$, and $T_E$) to the values listed in Table\,\ref{table:C_fit_para}.

We then estimate the magnetic entropy from the residual $C - C_\textrm{ph}$.
As shown in the inst of Fig.\,\ref{fig:C_SM}, the full entropy of $S=1/2$ is well recovered at $\sim 50$\,K which is close to $J/k_\textrm{B}=45$\,K~\cite{OkumaPRB2017}.
This good agreement validates this phenomenological phonon estimation.

\begin{table}[htb]
	\caption{Fit results}
	\label{table:C_fit_para}
	\begin{tabular}{ c  c  c  c } \toprule
		$A_D$ & $T_D$ & $A_E$ & $T_E$  \\ \midrule
		3.052 $\pm$ 0.052 & 219.8 $\pm$ 3.791 & 4.371 $\pm$ 0.286 & 840.3 $\pm$ 36.49 \\ \bottomrule
	\end{tabular}
\end{table}

\subsection{The different field dependence of $\kappa_{xy}$ in the AFM phase of Cd-K and Ca-K}

As shown in the main text, the field dependence of $\kappa_{xy}$ in the AFM phase of Cd-K is quite contrast to that of Ca-K~\cite{DokiPRL2018}, whereas the AFM order of both compounds is suggested to be the same $q = 0$ negative chiral state\,\cite{OkumaPRB2017, Ihara2020, Iida2020}.
 
In Cd-K, a finite $\kappa_{xy}$ is observed only above 7\,T. This field-induced $\kappa_{xy}$ is found to be the phonon thermal Hall $\kappa_{xy}^\textrm{ph}$ caused by the field-induced spin excitations. On the other hand, in Ca-K, $\kappa_{xy}$ is observed only below $\sim 6$\,T.

The absence of $\kappa_{xy}$ in Ca-K at high fields means the absence of  $\kappa_{xy}^\textrm{ph}$.
This is consistent with the small phonon thermal conductivity in Ca-K which is strongly suppressed by the crystal inhomogeneity owing to the random distribution of the Ca$^{2+}$ ions and the nonstoichiometric composition of H$_2$O molecules~\cite{YoshidaJPSJ2017}.
Moreover, the re-emergence of $\kappa_{xy}^\textrm{ph}$ in Cd-K is caused by the field-induced spin excitations through a spin-phonon coupling. On the other hand, such field-induced excitation is absent in the AFM phase of Ca-K. The different $\kappa_{xy}$ at high fields can be understood by these differences.

The absence of the phonon thermal Hall in Ca-K also means that $\kappa_{xy}$ at low fields in the AFM phase of Ca-K is brought by spins. On the other hand, $\kappa_{xy}^\textrm{sp}$ is absent in the AFM phase of Cd-K. This difference may be caused by a different spin rotation angle in the negative chiral state. 
According to the theoretical study~\cite{Mook2019}, $\kappa_{xy}^\textrm{sp}$ in the negative chiral state depends on the spin rotation angle in the kagome plane.
In Cd-K, the absence of $\kappa_{xy}^\textrm{sp}$ has been suggested owing to the local $\left\langle 100 \right\rangle $ anisotropy~\cite{OkumaPRB2017}, which is consistent with our experimental results.
Therefore, the finite $\kappa_{xy}^\textrm{sp}$ at low fields in Ca-K may suggest a different rotation angle of the spin in the negative chiral state. 
Also, the finite $\kappa_{xy}^\textrm{sp}$ in Ca-K could be attributed to an inhomogeneous magnetic state in Ca-K caused by the crystal inhomogeneity.
This is because $\kappa_{xy}^\textrm{sp}$ is affected by the magnetic domain structure owing to the spin-angle dependence.
In fact, the NMR measurement in Ca-K\,\cite{Ihara2020} reports a broadened NMR spectrum in the AFM state, whereas a two-peak spectrum is expected in the $q=0$ state. On the other hand, such inhomogeneity is absent in Cd-K, which is also consistent with the absence of $\kappa_{xy}^\textrm{sp}$ in the AFM state.

\end{document}